\documentclass{elsart}
\usepackage{epsfig,subfigure,latexsym,amsmath}
 
\begin{document}
 
\input epsf
 
\begin{frontmatter}
 
\title{Total Prompt Energy Release \\ in the Neutron-Induced Fission
\\ of $^{235}$U, $^{238}$U, and $^{239}$Pu}
\author{D. G. Madland}
\address{Theoretical Division, Los Alamos National Laboratory,
Los Alamos, New Mexico 87545}
 
\begin{abstract}
This study addresses, for the first time, the total prompt energy release and its components
for the fission of $^{235}$U, $^{238}$U, and $^{239}$Pu as a function of the kinetic energy
of the neutron inducing the fission. The components are extracted from experimental
measurements, where they exist, together with model-dependent calculation,
interpolation, and extrapolation. While the components display clear dependencies upon the
incident neutron energy, their sums display only weak, yet definite, energy dependencies.
Also addressed is the total prompt energy deposition
in fission for the same three systems. Results are presented in equation form.
New measurements are recommended as a consequence of this study.
\end{abstract}
 
\begin{keyword}
Energy release and energy deposition in neutron-induced fission, experiment and Los Alamos model,
$^{235}$U, $^{238}$U, $^{239}$Pu
\PACS 24.75.+i \sep 25.85.Ec \sep 25.85.Ca \sep 27.90.+b
\end{keyword}
\end{frontmatter}
 
\section{Introduction}
 
\noindent This study is a consequence of open questions on the magnitudes of the total
prompt energy release in fission, the total prompt energy deposition in fission,
the components of these quantities, and their
dependencies upon the kinetic energy of the neutron inducing the fission.
Our results are given in Eqs. (\ref{u5er} - \ref{pu9er}) and Fig. \ref{fig22} for the total
prompt energy release in fission, and in Eqs. (\ref{edu5} - \ref{edpu9})
and Fig. \ref{fig23} for the total prompt energy deposition in fission.
Recommended new experimental measurements coming
from this study are given in Sec. 6.
It should be noted that the study relies primarily upon existing published experimental
measurements and secondarily upon nuclear theory and nuclear models.
Therefore, new and higher quality measurements would improve the results presented
here and, furthermore, would lead to a more complete understanding
of post scission fission physics.
 
\section{Energy Conservation}
 
\noindent The energy release in fission is obtained from energy conservation. If one considers
the binary fission of an actinide nucleus of mass number $A-1$ induced by a neutron of mass
$m_{n}$, kinetic energy $E_{n}$, and binding energy $B_{n}$ in the compound nucleus $A$
formed when the neutron is absorbed, then energy conservation gives
 
\begin{eqnarray}
E_{n} + m_{n} + M(Z,A-1) &=& T  + M^{*}(Z,A ) \nonumber \\
&=& T_{L}(Z_{L},A_{L}) + M_{L}^{*}(Z_{L},A_{L}) \nonumber \\
&+& T_{H}(Z_{H},A_{H}) + M_{H}^{*}(Z_{H},A_{H})
\label{eq1}
\end{eqnarray}
 
\noindent where the left side specifies the initial conditions prior to fission, the right side
specifies the excited compound nucleus that is about to fission, and the second right
side specifies the excited fission fragments just after binary fission has occurred.
The notation here is that $T$ is a compound nucleus or fragment kinetic energy, $M$ and $m$ are stable masses, $M^{*}$
are masses of excited nuclei, all in units of MeV ($c^{2}$ has been suppressed),
and $L$ and $H$ refer to the light and heavy fragments occurring in the binary fission.
 
\noindent The neutron binding energy $B_{n}$ is obtained from the Q-value for neutron
capture,
 
\begin{equation}
m_{n} + M(Z,A-1) = M(Z,A) + B_{n}
\label{eq2}
\end{equation}
 
\noindent which, when inserted into Eq. (1), yields
 
\begin{eqnarray}
E_{n} + B_{n} + M(Z,A) &=& T_{L}(Z_{L},A_{L}) + T_{H}(Z_{H},A_{H}) + M_{L}(Z_{L},A_{L}) \nonumber \\
&+& \mbox{} M_{H}(Z_{H},A_{H}) + E_{L}^{*}(Z_{L},A_{L}) + E_{H}^{*}(Z_{H},A_{H})
\label{eq3}
\end{eqnarray}
 
\noindent where we have written an excited fragment mass as the ground-state mass plus the excitation
energy, namely, $M^{*} = M + E^{*}$. This excitation energy will be dissipated by the emission of prompt neutrons
and prompt gamma rays.
 
\noindent The {\it total energy release $E_{r}$ in binary fission is defined as the ground-state mass
of the compound nucleus undergoing fission minus the ground-state masses of
the two binary fission fragments}, namely,
 
\begin{equation}
E_{r} = M(Z,A) - M_{L}(Z_{L},A_{L}) - M_{H}(Z_{H},A_{H})
\label{eq4}
\end{equation}
 
\noindent where (again) the masses are expressed in units of MeV \cite{Halp}.
This equation also defines the total energy release in each stage of multiple-chance
fission except for the average kinetic and binding energies of the neutron(s)
emitted prior to fission in each stage, which must be taken into account.
Inserting Eq. (\ref{eq4}) into Eq. (\ref{eq3}) yields a second expression for the
energy release in fission:
 
\begin{equation}
E_{r} = T_{L}(Z_{L},A_{L}) + T_{H}(Z_{H},A_{H}) + E_{L}^{*}(Z_{L},A_{L}) + E_{H}^{*}(Z_{H},A_{H}) - (E_{n} + B_{n})
\label{eq5}
\end{equation}
 
\noindent In these equations, conservation of charge ensures that
$Z = Z_{L} + Z_{H}$, and conservation of baryon number ensures that
$A = A_{L} + A_{H}$. Note that in Eq.\ (\ref{eq5}) the
kinetic and binding energy of the neutron inducing fission explicitly appear,
with a minus sign, but they also implicitly appear in the fragment excitation
energies $E_{L}^{*}$ and $E_{H}^{*}$. And for spontaneous fission one replaces
($E_{n} + B_{n}$) with $0$, while for photofission the replacement is $E_{\gamma}$.
 
\noindent If we now sum the fragment kinetic energies and
fragment excitation energies, Eq.\ (\ref{eq5}) becomes
 
\begin{equation}
E_{r} = T_{f}^{tot} + E_{tot}^{*} - (E_{n} + B_{n})
\end{equation}
 
\noindent with
 
\begin{eqnarray}
T_{f}^{tot} &=& T_{L}(Z_{L},A_{L}) + T_{H}(Z_{H},A_{H}) \\
\label{eq7}
E_{tot}^{*} &=& E_{L}^{*}(Z_{L},A_{L}) + E_{H}^{*}(Z_{H},A_{H}) \;.
\label{eq8}
\end{eqnarray}
 
\noindent The above equations are for the specific binary fission $(Z_{L},A_{L}) +
(Z_{H},A_{H})$.
 
\noindent However, a large number of binary mass splits are energetically
allowed and they have been observed throughout the pre-actinide, actinide, and
trans-actinide regions \cite{Gonn}. In the vicinity of the uranium and plutonium isotopes
the fission-fragment mass range is (approximately) $70 \leq A_{f} \leq 170$ and
for each $A_{f}$ there are 4 to 5 contributing isobars leading to between
200 and 250 different possible mass splits in binary fission ($f$ stands for fragment).
This means that the total energy release $E_{r}$, the
total fission-fragment kinetic energy
$T_{f}^{tot}$, and the total fission-fragment excitation energy $E_{tot}^{*}$, in Eqs.\ (\ref{eq4}-\ref{eq8}),
must be replaced by their average values as determined by weighting with the
independent fission-fragment yields $Y_{f}$, where
 
\begin{equation}
Y_{f}(Z_{L},A_{L}) = Y_{f}(Z-Z_{L},A-A_{L}) = Y_{f}(Z_{H},A_{H})
\label{eq9}
\end{equation}
 
\noindent leading to
 
\begin{eqnarray}
\langle E_{r}\rangle &=& M(Z,A) - \frac{\sum{[Y_{f}] [M_{L}(Z_{L},A_{L}) + M_{H}(Z_{H},A_{H})]}}{\sum{Y_{f}}} \\
\label{eq10}
&=& \langle T_{f}^{tot}\rangle + \langle E_{tot}^{*}\rangle - (E_{n} + B_{n})
\label{eq11}
\end{eqnarray}
 
\noindent with
 
\begin{eqnarray}
\langle T_{f}^{tot}\rangle = \frac{\sum{[Y_{f}] [T_{L}(Z_{L},A_{L}) + T_{H}(Z_{H},A_{H})]}}{\sum{Y_{f}}} \\
\label{eq12}
\langle E_{tot}^{*}\rangle = \frac{\sum{[Y_{f}] [E_{L}^{*}(Z_{L},A_{L}) + E_{H}^{*}(Z_{H},A_{H})]}}{\sum{Y_{f}}}
\label{eq13}
\end{eqnarray}
 
\noindent and the sums are understood to be over either the light \{$L$\} or the
heavy \{$H$\} fission-fragment yields.
Thus, Eqs.\ (\ref{eq9}-\ref{eq13}) replace Eqs.\ (\ref{eq4}-\ref{eq8}) for the total energy release in binary fission.
Note, again, that in the case of spontaneous fission $E_{n}$ and $B_{n}$ in Eq.\ (\ref{eq11})
are set to zero, while in the case of photofission they are replaced with $E_{\gamma}$.
 
\noindent Extraction of the average total {\it prompt} energy release
in fission from Eqs.\ (\ref{eq9}-\ref{eq13})
requires consideration of the time dependence of the fission process and the introduction
of definitions related to that time dependence.
A schematic of neutron-induced binary fission is shown in Fig. \ref{fig1}. The
terms appearing in the figure, as well as others, are defined as follows:
 
\begin{figure} [htb]
\vspace{-75pt}
\begin{center}
\epsfig{file=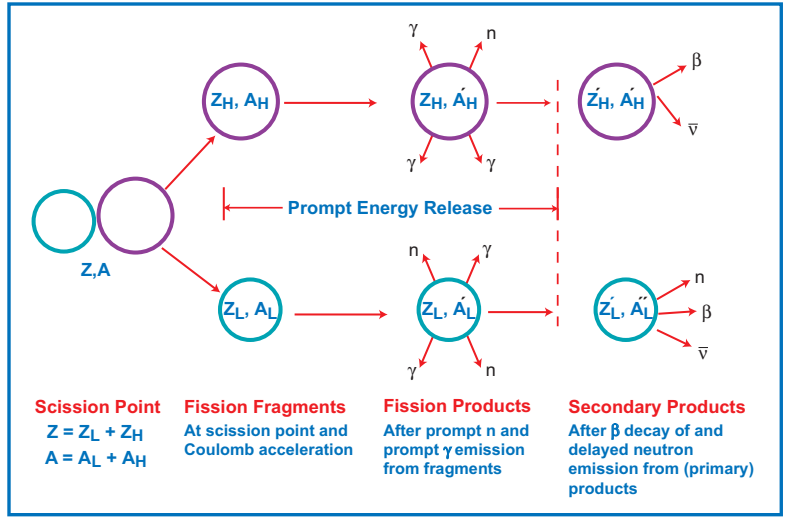, width=7.0in,angle=0.}
\end{center}
\vspace{-75pt}
\caption{Schematic of post scission in neutron-induced binary
fission of target nucleus $(Z,A-1)$.}
\label{fig1}
\end{figure}
 
\begin{description}
\item[{\it scission point:}] The time at which the fission fragments are unalterably determined \cite{Wil}.
Loosely, the time at which the neck snaps between the nascent fission fragments.
\item[{\it fission fragment:}] Nuclear species existing at the scission point and just beyond,
but prior to the emission of prompt neutrons and prompt gamma rays.
\item[{\it fission fragment acceleration time:}] $\sim 10^{-20}$ [sec] due to Coulomb repulsion.
\item[{\it prompt neutron emission time:}] In the range $\sim 10^{-18} $ to $ \sim 10^{-13}$ [sec] based upon measurement of compound nucleus lifetimes and calculation.
\item[{\it prompt gamma emission time:}] In the range $\sim 10^{-14} $ to $ \sim 10^{-7}$ [sec] based upon time-of-flight measurements and calculation.
See Ref. \cite{note}.
\item[{\it prompt energy release time:}] In the range $\sim 10^{-20}$ to $\sim 10^{-7}$ [sec].
See Ref. \cite{note}.
\item[{\it fission product}] (or {\it primary fission product}): Nuclear species existing following
prompt neutron emission and prompt gamma emission from a fragment, but
before any $\beta$ decay has occurred.
\item[{\it secondary fission product:}] Nuclear species existing following at least one $\beta$
decay of a primary fission product. The shortest known fission-product $\beta$ decay half-life is 0.032 [sec].
Therefore, secondary fission products, $\beta$-decay energy spectra, antineutrino energy spectra, and subsequent delayed neutron
energy spectra play no role in the total prompt fission energy release.
\end{description}
 
\noindent Now the independent fission-fragment yields $Y_{f}(Z_{f},A_{f})$ required
in the solution of Eqs.\ (\ref{eq9}-\ref{eq13}) can only be
obtained by construction from the measured independent fission-product yields $Y_{p}(Z_{p},A_{p})$
(where $p$ stands for product), the measured average prompt neutron multiplicity as a function of
fission-fragment mass ${\bar{\nu}}_{p}(A_{f})$ (where $p$ stands for prompt), and a Gaussian or Gaussian-like model assumption for the $Z_{f}$
dependence of $Y_{f}$ for fixed $A_{f}$. Note that $Z_{p} = Z_{f}$, but that $A_{p} \leq A_{f}$
due to the prompt neutron emission from the fragment.
Furthermore, the independent fission-product yields $Y_{p}(Z_{p},A_{p})$ have been extensively measured,
and tabulated, only for spontaneous fission and neutron-induced fission at two well-defined
energies, thermal and 14 MeV (these are the well-known double-humped mass yield $(A_{p})$ distributions
together with Gaussian charge yield $(Z_{p})$ distributions) \cite{ENDF}.
Thus, solution of Eqs.\ (\ref{eq9}-\ref{eq13}) by use of constructed independent fission-fragment yields
$Y_{f}(Z_{f},A_{f})$ is not currently tractable except for incident thermal neutrons and 14-MeV neutrons.
Therefore, we turn to direct use of measured and/or calculated values of the
terms appearing in Eq.\ (\ref{eq11}). Before proceeding we note the following:
 
\begin{description}
\item[{\it multiple-chance fission:}] Insofar as measured quantities are
used in the evaluation of Eq. (\ref{eq11}) the effects of multiple-chance
fission are automatically taken into account. However, as will be seen, some
calculated quantities are yet needed and these require multiple-chance
fission treatment [see Eqs. (\ref{epscm}) and (\ref{emlab})]. Appendix A
contains the multiple-chance fission equations to be used in the calculation
of the average total prompt fission energy release and energy deposition,
$\langle E_{r}\rangle$ and $\langle E_{d}\rangle$ respectively,
when none of the measured components of these quantities are available.
\item[{\it ternary fission:}] For the incident neutron energy range
$0 \leq E_{n} \leq 15$ MeV, approximately 1 in 500 fissions is
ternary \cite{WA}. Therefore, ternary fission has been ignored in the preceeding
equations. However, insofar as ternary fission events have affected the
measurements to be used below, their influence is present.
Here, {\it ternary} means light charged particle accompanied fission.
\item[{\it scission neutrons:}] The question of neutron emission at the
scission point remains an open one, with experimental results ranging
from 0\% to 10\% of the total average prompt neutron multiplicity
${\bar{\nu}}_{p}$ \cite{WA}. Measurements of this quantity, to be used
below, include the scission neutrons if they exist.
\item[{\it isomeric states:}] The de-excitation of fission fragments to (long-lived)
isomeric states has been ignored in the preceeding equations and time definitions.
However, their effects are included insofar as they affect the measurements to
be used below.
\end{description}
 
\section{Components of the Average Total Prompt Fission Energy Release}
 
\noindent The components of the average total prompt fission energy release $\langle E_{r}\rangle$ to be evaluated
for the solution of Eq. (\ref{eq11}) are the average total fission-fragment kinetic energy
$\langle T_{f}^{tot}\rangle$ and the average total fission-fragment excitation
energy $\langle E^{*}_{tot}\rangle$ whereas $E_{n}$ and $B_{n}$ are known.
The largest component
is the average total fission-fragment kinetic energy $\langle T_{f}^{tot}\rangle$ which becomes, after prompt neutron
emission times ranging from 10$^{-18}$ to 10$^{-13}$ [sec], the average total fission-product
kinetic energy $\langle T_{p}^{tot}\rangle$  which is the measured quantity.
More often than not, the experimentalists have converted the measured {\it product}
kinetic energies back to {\it fragment} kinetic energies because they are
more relevant to the development of fission theory. These quantities
are related by the kinetic energies of the prompt neutrons emitted from the moving
fragments as they become moving products.
As a function of the kinetic energy $E_{n}$ of the neutron inducing fission, one
obtains
 
\begin{equation}
\langle T_{p}^{tot}(E_{n})\rangle \;=\; \langle T_{f}^{tot}(E_{n})\rangle \left[1\,-\,\frac{\bar{\nu}_{p}(E_{n})}{2A}\,\left(\frac{\langle A_{H} \rangle}{\langle A_{L} \rangle}
\,+\,\frac{\langle A_{L} \rangle}{\langle A_{H} \rangle}\,\right)\,\right]
\label{eq15}
\end{equation}
 
\noindent Equation (\ref{eq15}) yields about a 2\% kinetic energy correction due to prompt neutron emission
from fully accelerated fragments coming from 14-MeV neutron-induced fission.
The approximations used in its derivation are:
\begin{description}
\item [a) ] $m_{n}/M(Z,A)\;=\;1/A$
\item [b) ] ${\bar{\nu}}_{p}(L)\;=\;{\bar{\nu}}_{p}(H)\;=\;{\bar{\nu}}_{p}/2$ where
$L$ and $H$ refer to the average light and average heavy fragments
\item [c) ] $T_{f}/A\;=\;T_{f1}/(A-1)\;=\;T_{f2}/(A-2)\;=\;...$ where
$T_{f}$ is the initial fragment kinetic energy of the initial fragment $A$,
and $T_{fi}$ are fragment kinetic energies following the $i\,$th neutron evaporation
from the moving fragment.
\end{description}
 
\noindent These approximations are sufficiently accurate to quantify
a correction of order 2\%.
 
\noindent The average total fission-fragment excitation energy $\langle E^{*}_{tot}\rangle$
is dissipated by two separate mechanisms: prompt neutron emission and prompt gamma emission,
where {\it prompt} time has already been specified for each mechanism. Thus,
 
\begin{equation}
\langle E^{*}_{tot}\rangle = \langle Ex^{tot}_{n}\rangle + \langle E^{tot}_{\gamma}\rangle
\label{exc}
\end{equation}
 
\noindent where the average fission-fragment excitation energy leading to
prompt neutron emission is given by \cite{MN}
 
\begin{equation}
\langle Ex^{tot}_{n}\rangle = {\bar{\nu}}_{p}[\langle S_{n}\rangle + \langle \varepsilon\rangle]
\label{excn}
\end{equation}
 
\noindent with $\langle S_{n}\rangle$ the average fission-fragment neutron separation energy
and $\langle \varepsilon\rangle$ the average center-of-mass energy of the emitted neutrons,
and the average fission-fragment excitation energy leading to prompt gamma emission
is given by $\langle E^{tot}_{\gamma}\rangle$.
Equation (\ref{eq11}) now becomes
 
\begin{eqnarray}
\langle E_{r}\rangle &=& \langle T_{f}^{tot}\rangle + \langle Ex^{tot}_{n}\rangle + \langle E^{tot}_{\gamma}
\rangle - (E_{n} + B_{n}) \nonumber \\
&=& \langle T_{f}^{tot}\rangle + {\bar{\nu}}_{p}[\langle S_{n}\rangle + \langle \varepsilon\rangle]
+ \langle E_{\gamma}^{tot}\rangle - (E_{n} + B_{n})
\label{eqer}
\end{eqnarray}
 
\noindent Note that all averaged quantities appearing in Eqs.\ (\ref{exc}-\ref{eqer}) depend upon
the incident neutron energy $E_{n}$ which has been suppressed for brevity.
We use Eq.\ (\ref{eqer}) for the average total prompt energy release in
neutron-induced fission for the remainder of this paper.
 
\noindent The existing experimental database for the neutron-induced fission of $^{235}$U, $^{238}$U,
and $^{239}$Pu, together with model-dependent interpolation, extrapolation, and calculation,
allow a determination of $\langle E_{r}\rangle$ over the incident neutron energy range of
$0 \leq E_{n} \leq 15$ MeV. However, as will be seen below, the experimental database is
astonishingly incomplete.
Where experiment does exist,
we have performed linear or quadratic least-squares fits in $E_{n}$ to the data and
present the resulting parameters and their standard deviations in the following.
Standard (theoretical) deviations in the Los Alamos model, as used in the following,
are not quantitatively addressed herein.
 
\subsection{Average Total Fission Fragment and Fission Product Kinetic Energy}
 
\begin{figure} [htb]
\vspace{-0.5in}
\epsfig{file=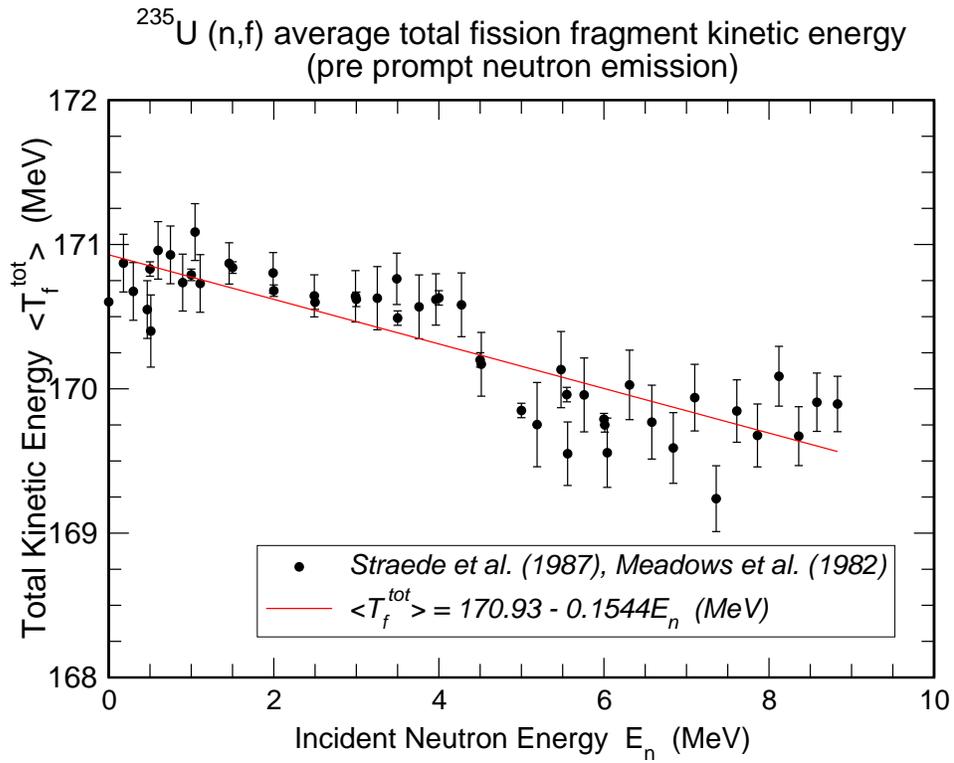, width=3.0in,angle=-90.}
\vspace{90pt}
\caption{Average total fission-fragment kinetic energy for the n(E$_{\rm n}$) + $^{235}$U system.}
\label{fig2}
\end{figure}

\begin{figure} [b!]
\vspace{-0.5in}
\epsfig{file=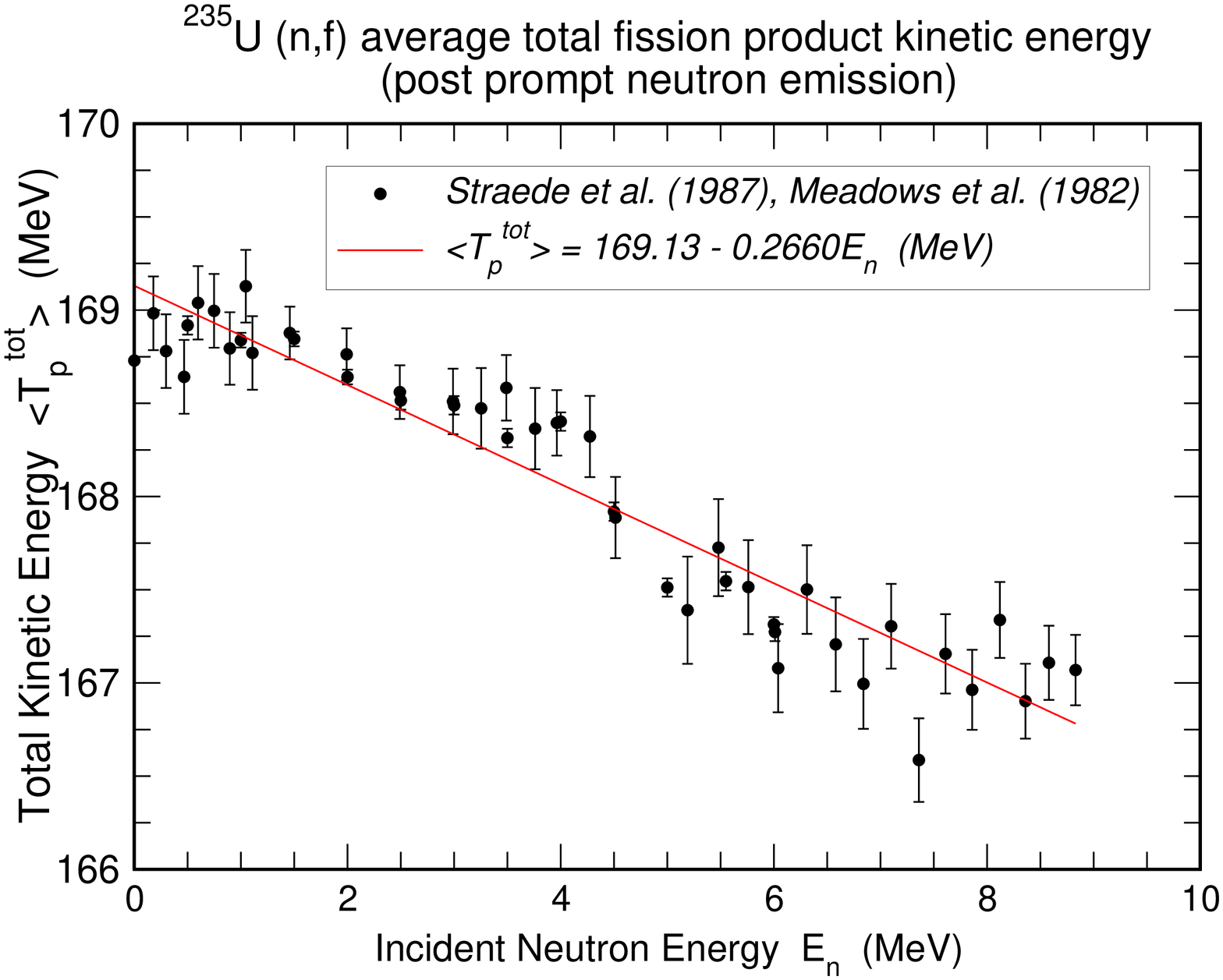, width=3.0in,angle=-90.}
\vspace{90pt}
\caption{Average total fission-product kinetic energy for the  n(E$_{\rm n}$) + $^{235}$U  system.}
\label{fig3}
\end{figure}

\noindent The experimental data that we use for the n + $^{235}$U system are
those of Meadows and Budtz-Jorgensen (1982) \cite{Me82}, Straede {\it et al.}
(1987) \cite{ST87}, and M\"uller {\it et al.} (1984, two data points only)
\cite{Mu84}. The published fission-fragment (pre prompt neutron emission)
total kinetic energies are shown in Fig. \ref{fig2} and the corresponding
fission-product (post prompt neutron emission) total kinetic energies,
obtained with Eq. (\ref{eq15}), are shown in Fig. \ref{fig3}. Linear fits
to these data are also shown in the figures.
 
\noindent {\bf For the n + $^{235}$U system:}
 
\begin{eqnarray}
\langle T_{f}^{tot} \rangle &=& (170.93 \pm 0.07) - (0.1544 \pm 0.02)E_{n} \; \;  (MeV) \label{eq16} \\
\langle T_{p}^{tot} \rangle &=& (169.13 \pm 0.07) - (0.2660 \pm 0.02)E_{n} \; \;  (MeV) \; .
\label{eq17}
\end{eqnarray}
 
\noindent The data appear to have structure (near the second-chance fission
threshold, for example), but their scatter and uncertainties preclude anything
other than a linear fit. The steeper negative slope for the total fission-product
kinetic energy is due to the energy dependence of ${\bar{\nu}}_{p}$ in Eq. (\ref{eq15}).
Strictly, Eqs.\ (\ref{eq16}) and (\ref{eq17}) should not be used above an incident neutron
energy of about 9 MeV.
 
\begin{figure} [htb]
\vspace{-0.5in}
\epsfig{file=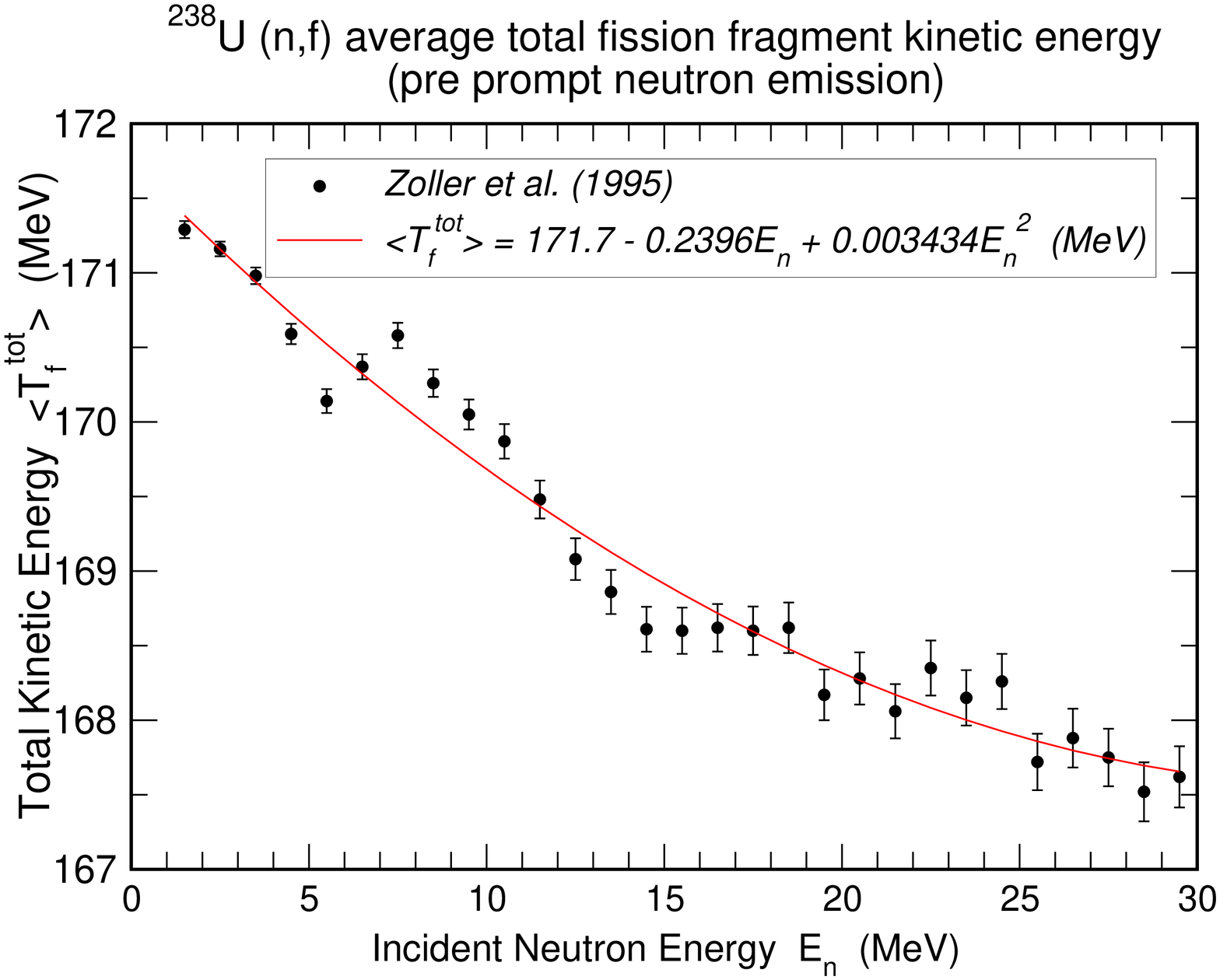, width=3.0in, angle=-90.}
\vspace{90pt}
\caption{Average total fission-fragment kinetic energy for the n(E$_{\rm n}$)
+ $^{238}$U system.}
\label{fig4}
\end{figure}

\begin{figure} [b!]
\vspace{-0.5in}
\epsfig{file=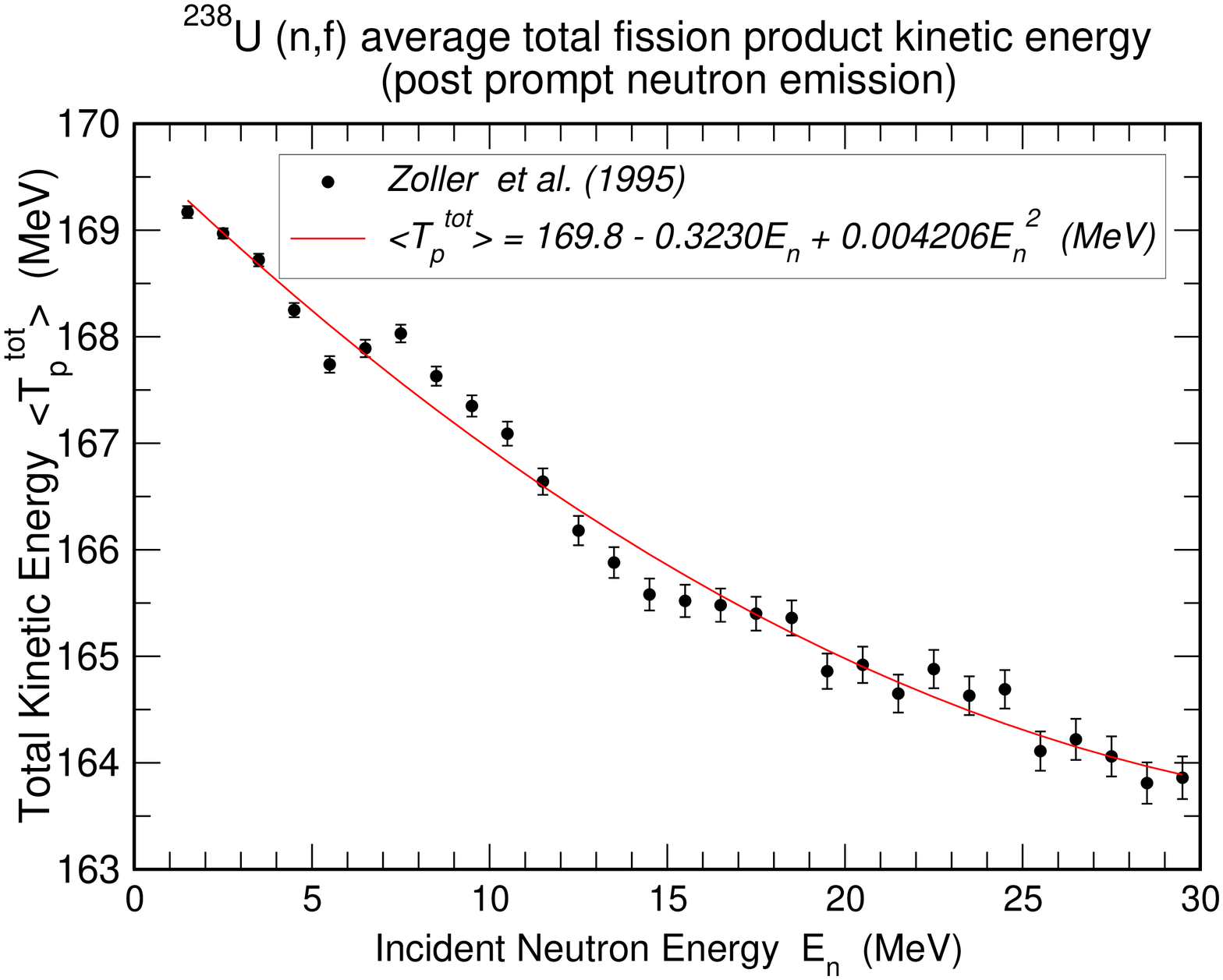, width=3.0in, angle=-90.}
\vspace{90pt}
\caption{Average total fission-product kinetic energy for the n(E$_{\rm n}$)
+ $^{238}$U system.}
\label{fig5}
\end{figure}

\noindent The experimental data that we use for the n + $^{238}$U system are
those of Z\"oller (1995) \cite{Zo95}
shown in Fig. \ref{fig4} and Fig. \ref{fig5} for incident neutron energies up
to 30 MeV. Here, the data give clear and convincing evidence for the presence of structure
near the second- and third-chance fission thresholds. For our present purposes, however,
we represent these data with quadratic fits for both the total fission-fragment
and total fission-product kinetic energies
because the corresponding experimental data for $^{235}$U and $^{239}$Pu are much lower in
quality and over more limited energy ranges.
We note that the recent experimental data of Viv$\grave{e}$s {\it et al.} (2000) \cite{Viv}
for this system,
over an incident neutron energy range of 1.2--5.8 MeV,
are in substantial agreement with the corresponding data of Z\"oller (Fig. \ref{fig4}).
The maximum discrepancy between the two measurements is $\sim$ 0.7\% at about 1.5 MeV.
 
\newpage
 
\newpage
 
\newpage
 
\begin{figure} [htb]
\vspace{-0.5in}
\epsfig{file=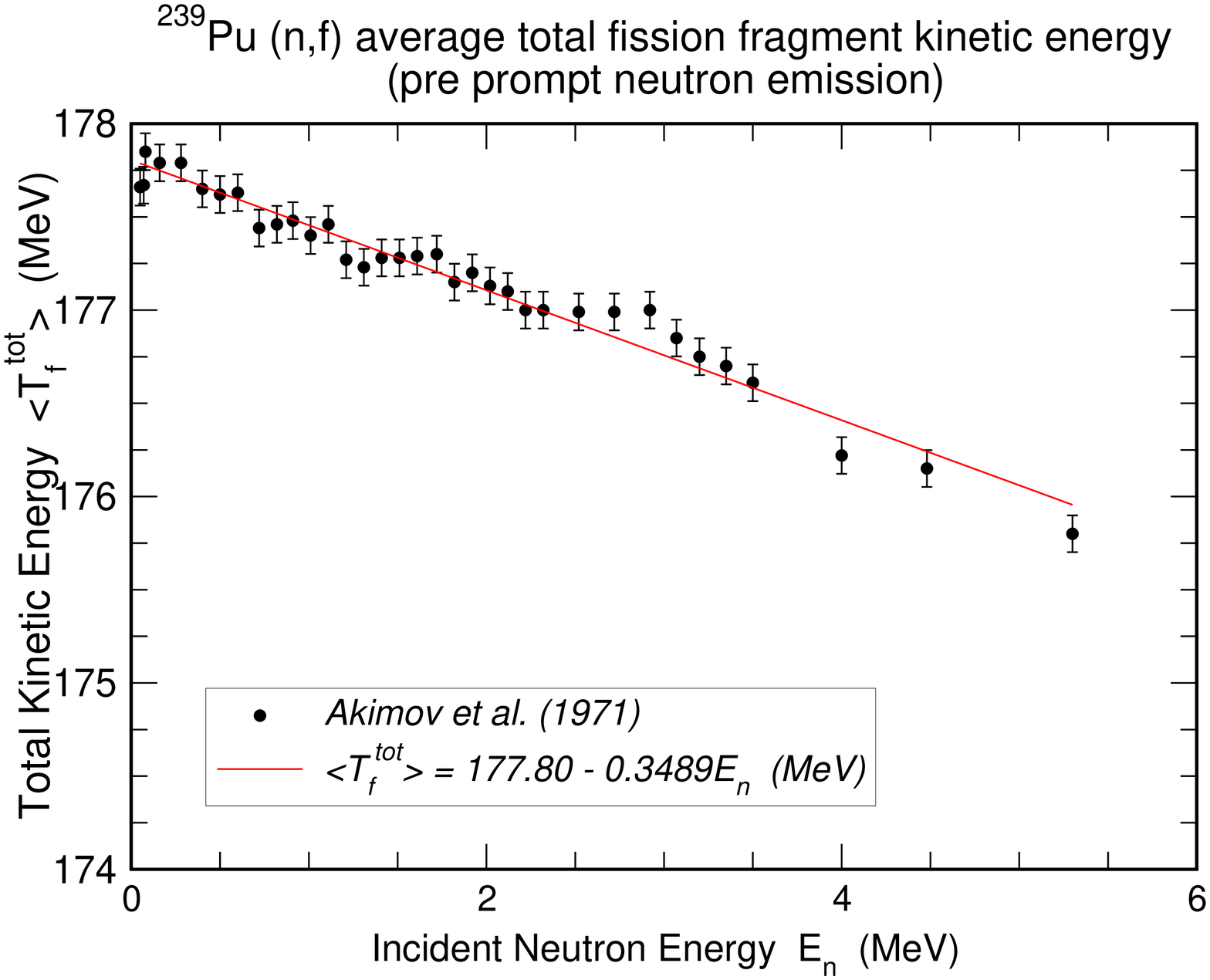, width=3.0in, angle=-90.}
\vspace{90pt}
\caption{Average total fission-fragment kinetic energy for the n(E$_{\rm n}$)
+ $^{239}$Pu system.}
\label{fig6}
\end{figure}
 
\begin{figure} [b!]
\vspace{-0.5in}
\epsfig{file=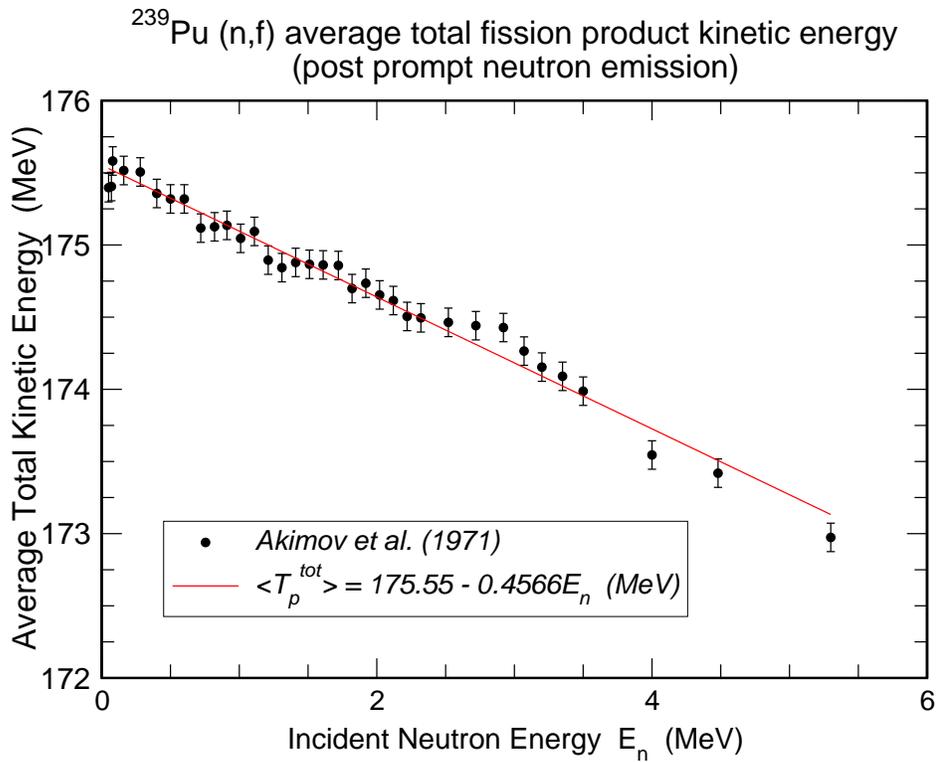, width=3.0in, angle=-90.}
\vspace{90pt}
\caption{Average total fission-product kinetic energy for the n(E$_{\rm n}$)
+ $^{239}$Pu system.}
\label{fig7}
\end{figure}
 
\newpage
 
\noindent {\bf For the n + $^{238}$U system:}
 
\begin{eqnarray}
\langle T_{f}^{tot} \rangle &=& (171.7 \pm 0.05) - (0.2396 \pm 0.01)E_{n} \nonumber \\
&+& \mbox{} (0.003434 \pm 0.0004)E_{n}^{2} \; \; (MeV)
\label{eq18} \\
\langle T_{p}^{tot} \rangle &=& (169.8 \pm 0.05) - (0.3230 \pm 0.01)E_{n} \nonumber \\
&+& \mbox{} (0.004206 \pm 0.0004)E_{n}^{2} \; \; (MeV)
\label{eq19}
\end{eqnarray}
 
\noindent The experimental data that we use for the n + $^{239}$Pu system
are those of Akimov {\it et al.} (1971) \cite{Ak71} shown in Fig. \ref{fig6} and
Fig. \ref{fig7} for incident neutron energies up to 5.5 MeV. Here, the data are
not high enough in incident neutron energy to ask whether structure exists
near the second-chance fission threshold.
 
\noindent Linear fits appear to be quite adequate
for the limited energy range and, strictly, the fits should not be used above
about 5.5 MeV. Note that in this system we have the steepest drop in the total
kinetic energies with increasing incident neutron energy of the three systems
under consideration.
 
\noindent {\bf For the n + $^{239}$Pu system:}
 
\begin{eqnarray}
\langle T_{f}^{tot} \rangle &=& (177.80 \pm 0.03) - (0.3489 \pm 0.02)E_{n} \; \; (MeV)
\label{eq20} \\
\langle T_{p}^{tot} \rangle &=& (175.55 \pm 0.03) - (0.4566 \pm 0.02)E_{n} \; \; (MeV) \; .
\label{eq21}
\end{eqnarray}
 
\noindent For all three of these systems the average total fission-fragment and fission-product
kinetic energies decrease with increasing incident neutron energy. The reason for this is that
the fission-fragment yields for symmetric and near-symmetric fission are increasing with
increasing incident neutron energy \cite{Gonn}, but the total kinetic energies are at, or near, a minimum
for symmetric and near-symmetric fission, thus decreasing the total kinetic energies
with increasing incident neutron energy. This effect has been observed in experiment.
See, for example, Fig. 11 of Ref. \cite{Sc66} for the n(thermal) + $^{235}$U system and
Fig. 4 of Ref. \cite{Ne66} for the n(thermal) + $^{239}$Pu system.
A (speculative) underlying physics reason for the observed effect may be that the
charge centers of the nascent fission fragments are slightly farther apart for
symmetric fission than they are for asymmetric fission.
 
\noindent The n + $^{239}$Pu system, Z=94, has the largest kinetic energies of the three
systems under study due to the Coulomb force, while those of the two U systems, Z = 92, are
comparable to each other and somewhat less.
 
\noindent The fission product kinetic energies, in addition to being the {\it largest}
component of the average total prompt energy release in fission, are the most
{\it localized} in energy deposition having corresponding ranges, for example, of $\sim 5 - 10$
[microns] in uranium.
 
\subsection{Average Total Prompt Neutron Emission Energy}
 
\noindent The average total prompt neutron emission energy is equal to
the average fission-fragment excitation energy leading to prompt
neutron emission $\langle Ex_{n}^{tot}\rangle$ as given by Eq.\ (\ref{excn}).
It is important to note that this quantity is not equal to the
average total prompt fission neutron kinetic energy $\langle E_{neut}^{tot}\rangle$
(to be discussed in Sec. 5) because (a) the portion of the prompt neutron kinetic energy
due to the motion of the fission fragments emitting the neutrons has not yet been
included and (b) the binding energy $S_{n}$ of the neutron emitted does not
contribute to the neutron kinetic energy.
 
\begin{figure} [htb]
\vspace{-0.5in}
\epsfig{file=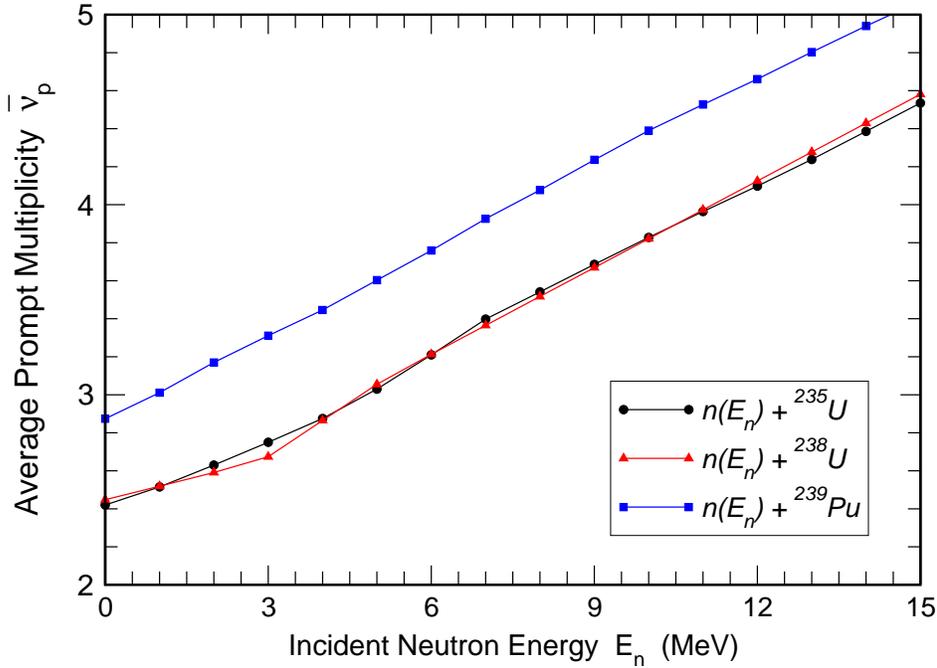, width=2.27in, angle=-90.}
\vspace{130pt}
\caption{Average prompt fission neutron multiplicity ${\bar{\nu}}_{p}$
for three systems from ENDF evaluated data (line segments are to guide the eye).}
\label{fig8}
\end{figure}

\noindent We evaluate Eq.\ (\ref{excn}) as a function of the incident neutron
energy $E_{n}$ as follows: First, the average prompt neutron multiplicities ${\bar{\nu}}_{p}(E_{n})$
are taken from the ENDF evaluations \cite{ENDF} which are based upon experimental data.
These are shown for the three systems under study in Fig. \ref{fig8}.
Second, the average fission-fragment neutron separation energy $\langle S_{n}\rangle$ is calculated
as one-fourth of the sum of the two two-neutron separation energies for a given binary
mass split in a seven-point approximation to the light and heavy mass peaks, as
described in Ref. \cite{MN}, so as to average over two each of the four possible odd-particle
configurations. We find $\langle S_{n}\rangle$ = 4.998 MeV for the n + $^{235}$U system,
$\langle S_{n}\rangle$ = 4.915 MeV for the n + $^{238}$U system, and $\langle S_{n}\rangle$
= 5.375 MeV for the n + $^{239}$Pu system.
For the range of incident neutron energies considered here, it is a reasonable
approximation that the $\langle S_{n}\rangle$ values are constant, independent
of the incident energy.
Third, the average center-of-mass energies $\langle \varepsilon\rangle$ of the emitted prompt neutrons
are calculated with the Los Alamos model \cite{MN}:
 
\begin{equation}
\langle \varepsilon\rangle = \frac{[P_{f_{1}}^{A}{\bar{\nu}}_{p_{1}}\langle \varepsilon_{1}\rangle +
P_{f_{2}}^{A}(\langle \xi_{1}\rangle + {\bar{\nu}}_{p_{2}}\langle \varepsilon_{2}\rangle)
+ P_{f_{3}}^{A}(\langle \xi_{1}\rangle + \langle \xi_{2}\rangle +
{\bar{\nu}}_{p_{3}}\langle \varepsilon_{3}\rangle)]}
{[P_{f_{1}}^{A}{\bar{\nu}}_{p_{1}} + P_{f_{2}}^{A}(1 + {\bar{\nu}}_{p_{2}}) +
P_{f_{3}}^{A}(2 + {\bar{\nu}}_{p_{3}})]}
\label{epscm}
\end{equation}
 
\noindent where, again, $A$ is the mass number of the fissioning compound nucleus,
the $P_{f_{i}}^{A}$ are the fission probabilities for $i$th-chance fission, the
${\bar{\nu}}_{p_{i}}$ are the average prompt neutron multiplicities for $i$th-chance
fission, the $\xi_{i}$ are the average kinetic energies of the evaporated neutrons
prior to fission in 2nd-chance fission ($i$ = 1) and 3rd-chance fission ($i$ = 2),
and the $\langle \varepsilon_{i}\rangle$ are the average center-of-mass neutron energies for
$i$th-chance fission.
 
\begin{figure} [b!]
\vspace{-0.5in}
\epsfig{file=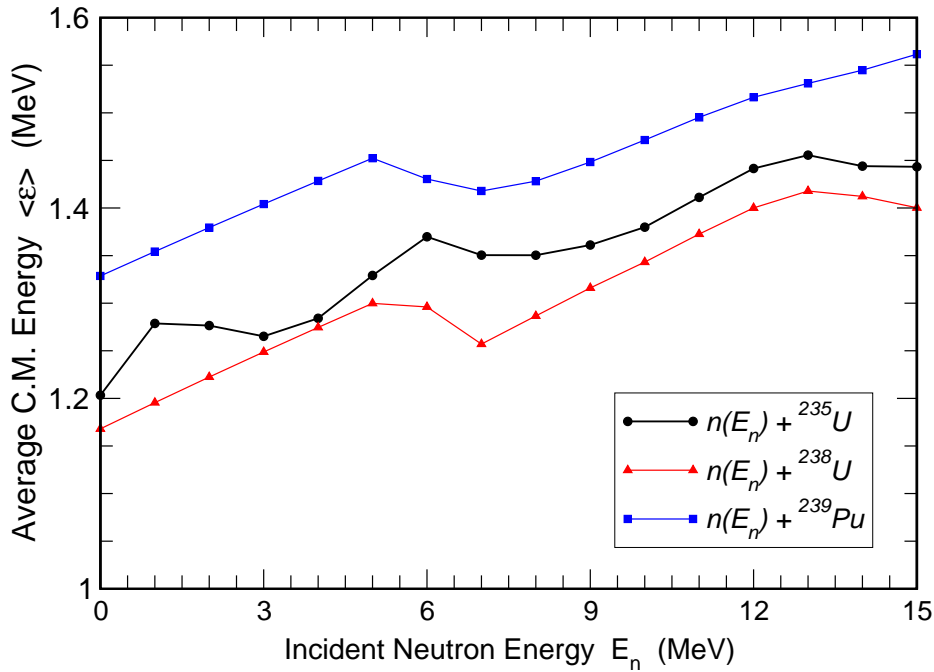, width=2.27in, angle=-90.}
\vspace{130pt}
\caption{Prompt fission neutron spectrum average center-of-mass energy $\langle \varepsilon\rangle$
for three systems calculated with the Los Alamos model (line segments
are to guide the eye).}
\label{fig9}
\end{figure}

\noindent Equation (\ref{epscm}) has been evaluated as a function of incident neutron energy for the three
systems under study and the results are shown in Fig. \ref{fig9}.
One sees that the values of $\langle \varepsilon\rangle$ generally increase with increasing
$E_{n}$, but decrease near 6 MeV and 13 MeV which are the approximate thresholds for
2nd- and 3rd-chance fission where the emission of 1 and 2 neutrons, respectively,
prior to fission, reduce the fragment excitation energy available for neutron and
gamma emission and, correspondingly, reduce $\langle \varepsilon\rangle$.
 
\noindent Note that
the structure observed below $\sim$ 3 MeV in the n + $^{235}$U system is due to fits
of the Los Alamos model to experimental spectra measured at these energies. It is
clear from Figs. \ref{fig8} and \ref{fig9} that the n + $^{239}$Pu system is the
{\it hottest} of the three systems in terms of both the number of neutrons emitted
and their energy, with the n + $^{235}$U and n + $^{238}$U systems very similar,
but somewhat larger neutron emission energies for n + $^{235}$U.
 
\noindent Using the preceding information,
Eq. (\ref{excn}) for the average fission-fragment excitation energy
leading to prompt neutron emission (average total prompt neutron emission energy)
can be evaluated for the three systems under study.
 
\begin{figure} [htb]
\vspace{-0.5in}
\epsfig{file=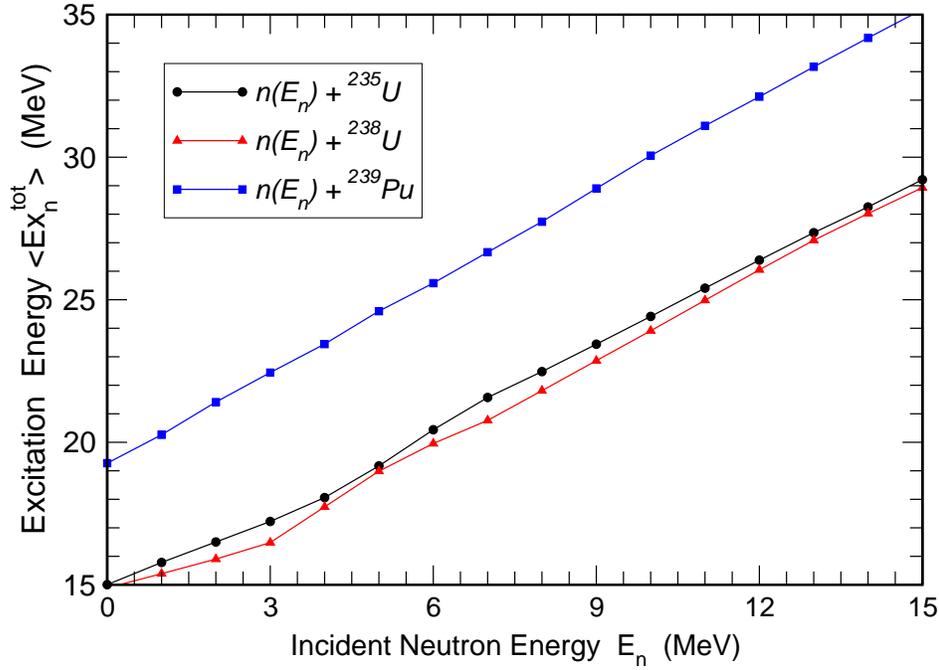, width=2.27in,angle=-90.}
\vspace{130pt}
\caption{Fission fragment excitation energy $\langle Ex_{n}^{tot}\rangle$ leading to prompt neutron emission
for three systems from the Los Alamos model and ENDF (line segments are
to guide the eye).}
\label{fig10}
\end{figure}

The results are shown in
Fig. \ref{fig10} which indicate an approximately linear dependence upon the incident
neutron energy $E_{n}$ and, again, more neutron emission for the Pu system and
less, but comparable, emission for the two U systems.
 
\begin{figure} [b!]
\vspace{-0.5in}
\epsfig{file=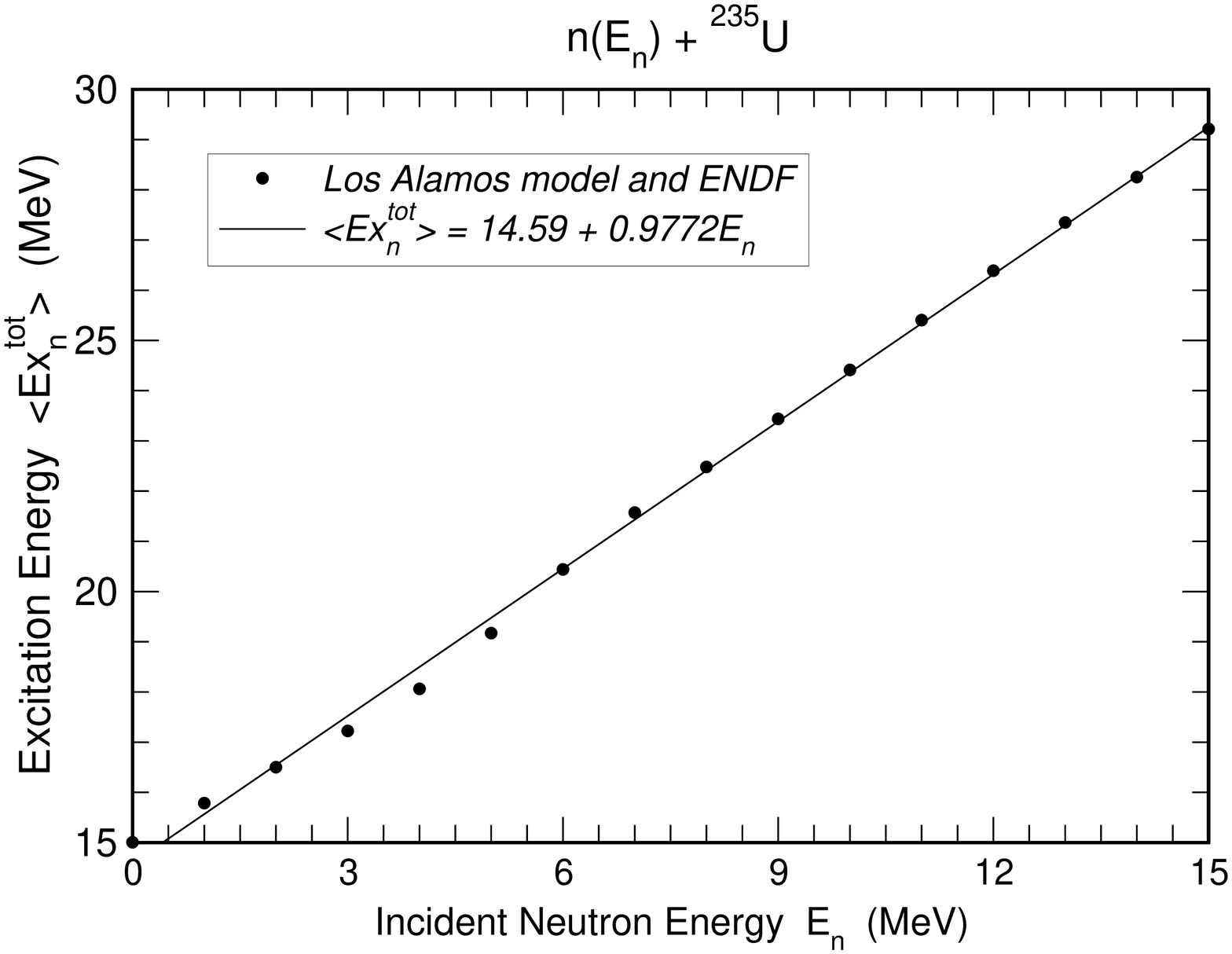,width=2.27in,angle=-90.}
\vspace{130pt}
\caption{Fission fragment excitation energy $\langle Ex_{n}^{tot}\rangle$ leading to prompt
neutron emission in the n(E$_{\rm n}$) + $^{235}$U system.}
\label{fig11}
\end{figure}

\begin{figure} [htb]
\vspace{-0.5in}
\epsfig{file=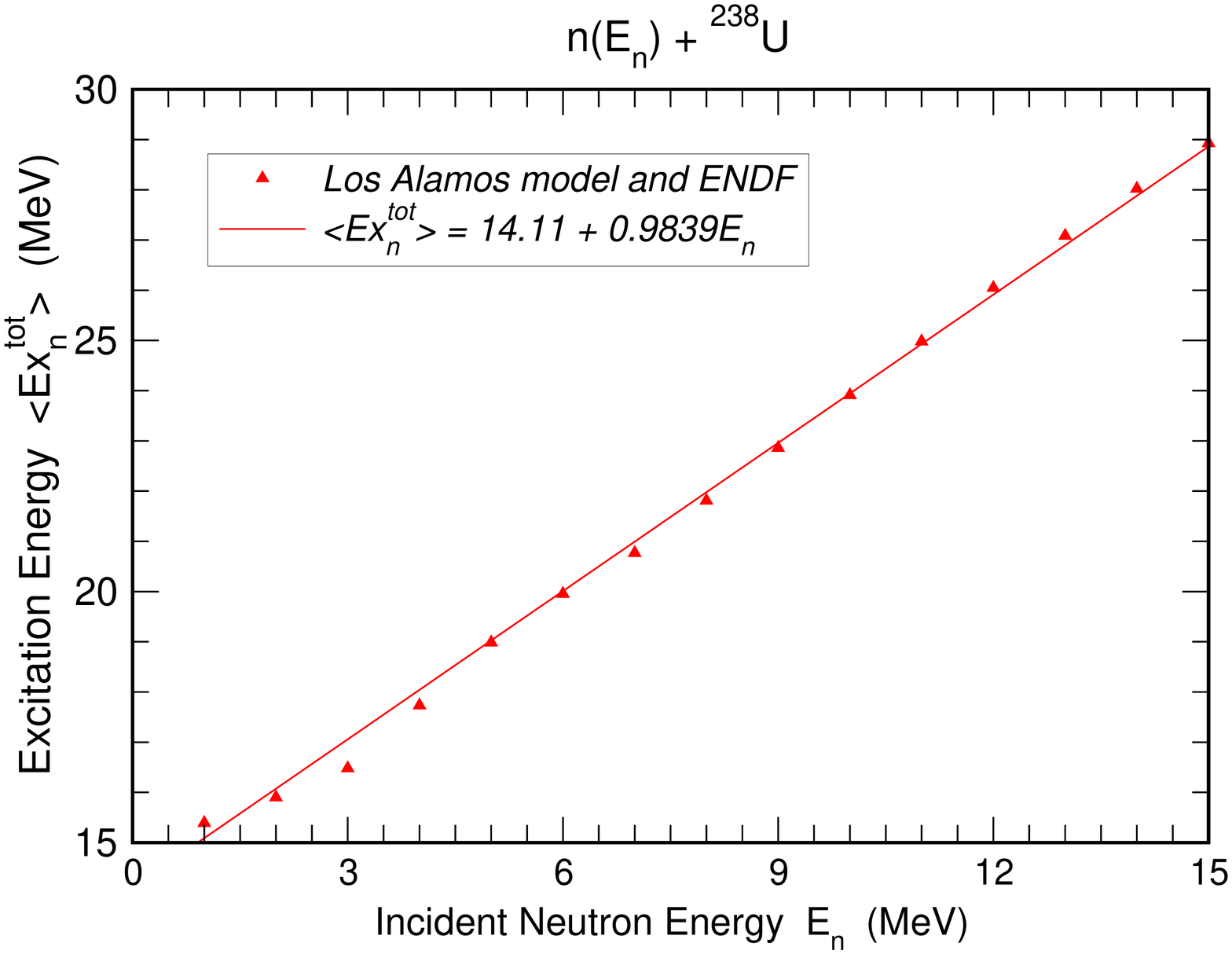,width=2.27in,angle=-90.}
\vspace{130pt}
\caption{Fission fragment excitation energy $\langle Ex_{n}^{tot}\rangle$ leading to prompt
neutron emission in the n(E$_{\rm n}$) + $^{238}$U system.}
\label{fig12}
\end{figure}

\begin{figure} [b!]
\vspace{-0.5in}
\epsfig{file=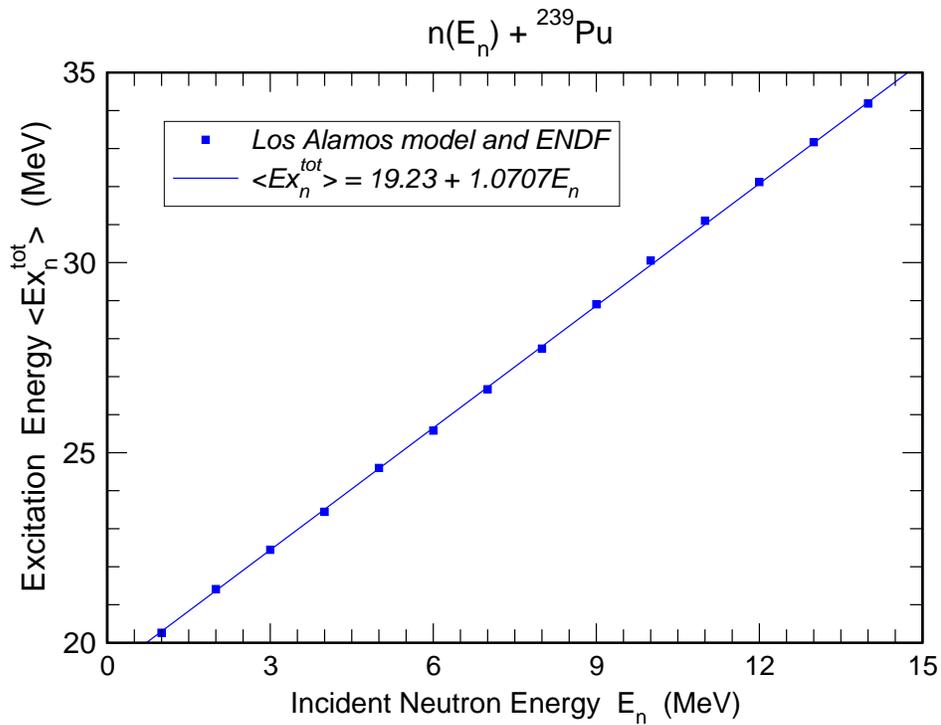,width=2.27in,angle=-90.}
\vspace{130pt}
\caption{Fission fragment excitation energy $\langle Ex_{n}^{tot}\rangle$ leading to prompt
neutron emission in the n(E$_{\rm n}$) + $^{239}$Pu system.}
\label{fig13}
\end{figure}

\noindent Linear fits to the three calculations shown in Fig. \ref{fig10} are
illustrated in Figs. \ref{fig11} - \ref{fig13} together with the calculated points
from Eq. (\ref{excn}) wherein the values of ${\bar{\nu}}_{p}(E_{n})$ are taken from ENDF. \\
 
\noindent {\bf For the n + $^{235}$U system:}
 
\begin{equation}
\langle Ex_{n}^{tot}\rangle = 14.59 + 0.9772E_{n} \; \; (MeV)
\label{u5ex}
\end{equation}
 
\noindent {\bf For the n + $^{238}$U system:}
 
\begin{equation}
\langle Ex_{n}^{tot}\rangle = 14.11 + 0.9839E_{n} \; \; (MeV)
\label{u8ex}
\end{equation}
 
\noindent {\bf For the n + $^{239}$Pu system:}
 
\begin{equation}
\langle Ex_{n}^{tot}\rangle = 19.23 + 1.0707E_{n} \; \; (MeV)
\label{pu9ex}
\end{equation}
 
\newpage
 
\newpage
 
\newpage
 
\newpage
 
\subsection{Average Total Prompt Gamma Emission Energy}
 
\begin{figure} [htb]
\vspace{-0.5in}
\epsfig{file=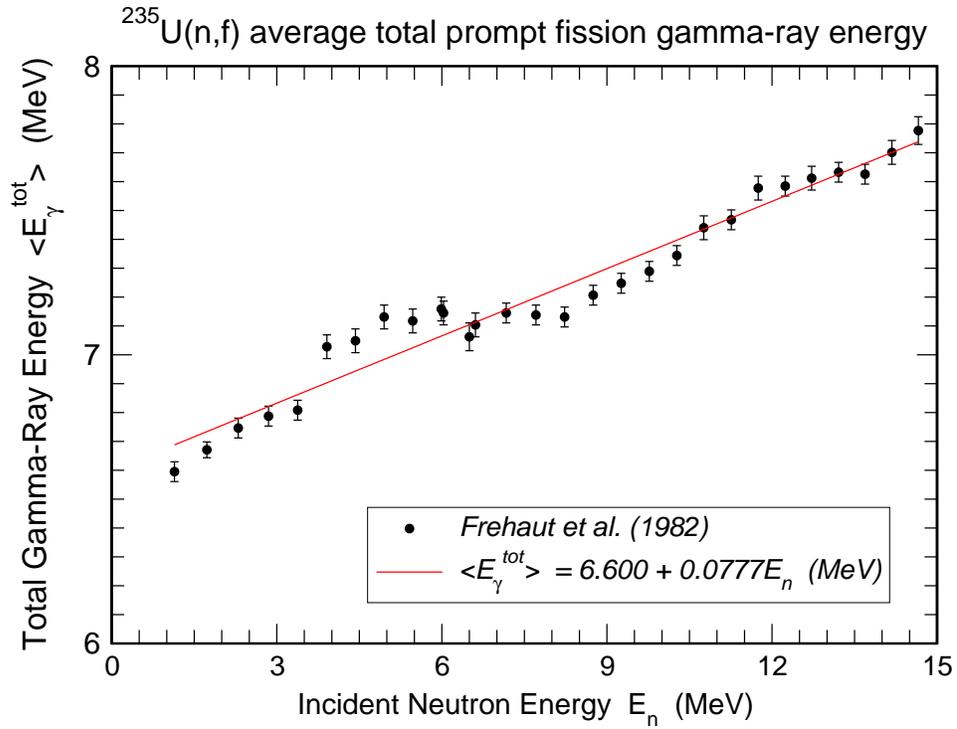, width=3.0in, angle=-90.}
\vspace{90pt}
\caption{Average total prompt fission gamma-ray energy
 $\rm \langle E_{\gamma}^{\rm tot}\rangle$ for the $\rm n(E_{\rm n}$) + $^{235}$U system.}
\label{fig14}
\end{figure}

\noindent The experimental data that we use for the n + $^{235}$U system are those of
Frehaut {\it et al.} (1982) \cite{Fr82} which are ratio measurements to the average
total prompt gamma emission energy for $^{252}$Cf(sf). These ratio data
have been converted back to absolute units by using the average of three measurements
for $^{252}$Cf reported in the Hoffman and Hoffman review \cite{HH74}.
The converted Frehaut {\it et al.} data are shown in Fig. \ref{fig14} together
with a linear fit to the data.
There is nonlinear structure in these data, but for our present purposes
we use the linear approximation.
 
\noindent {\bf For the n + $^{235}$U system:}
 
\begin{equation}
\langle E_{\gamma}^{tot}\rangle = (6.600 \pm 0.03) + (0.0777 \pm 0.004)E_{n} \; \; (MeV)
\label{u5gam}
\end{equation}

\begin{figure} [b!]
\vspace{-0.5in}
\epsfig{file=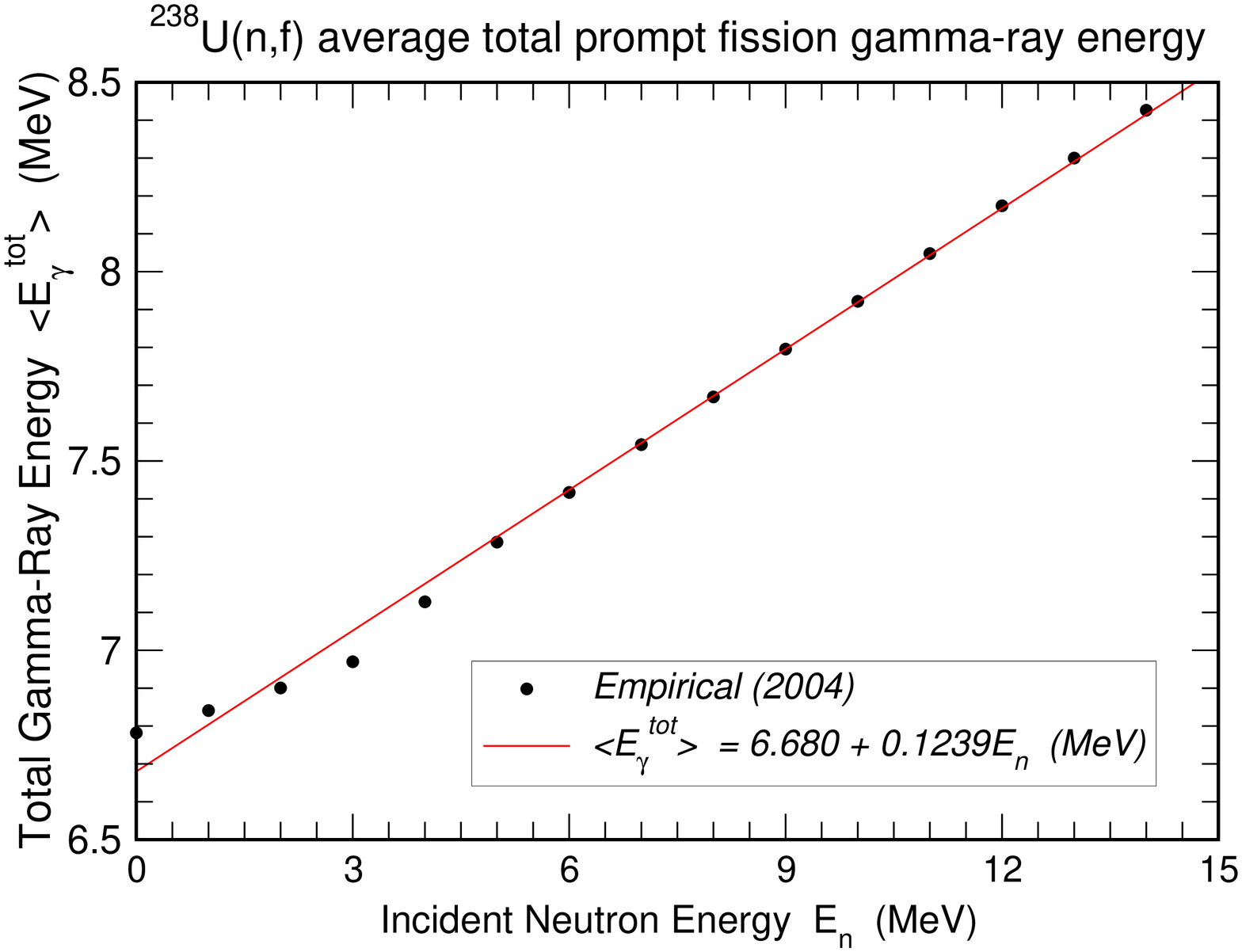, width=3.0in, angle=-90.}
\vspace{90pt}
\caption{Average total prompt fission gamma-ray energy
 $\rm \langle E_{\gamma}^{\rm tot}\rangle$ for the $\rm n(E_{\rm n}$) + $^{238}$U system.}
\label{fig15}
\end{figure}

\noindent We have no experimental values for the average total prompt gamma emission
energy in the n + $^{238}$U system. Therefore, an empirical approach is used, namely,
a linear assumption with ${\bar{\nu}}_{p}(E_{n})$ is made (based upon the Frehaut {\it et al.} \cite{Fr82} measurements) with
the zero (thermal) energy value taken from the $A$-dependent fit by Hoffman and
Hoffman \cite{HH74}, and the slope
taken from that inferred by Frehaut {\it et al.} from the n + $^{237}$Np measurements
that they performed. Note, however, that we use the lower experimental limit of
their inferred slope because the n + $^{237}$Np system is {\it hotter} than that
of n + $^{238}$U.
The resulting data points are labeled ``Empirical (2004)'' in Fig. \ref{fig15}
together with a linear fit in incident neutron energy.
 
\noindent {\bf For the n + $^{238}$U system:}
 
\begin{equation}
\langle E_{\gamma}^{tot}\rangle = 6.6800 + 0.1239E_{n} \; \; (MeV)
\label{u8gam}
\end{equation}
 
\noindent Experimental values exist for the average total prompt gamma emission energy
for the n + $^{239}$Pu system, but only for thermal neutron energy:
Pleasonton (1973) \cite{Pl73} measured a value of 6.73 $\pm$ 0.35 MeV for
the thermal case. Direct measurements for greater neutron energy do not appear
to exist.
 
\begin{figure} [htb]
\vspace{-0.5in}
\epsfig{file=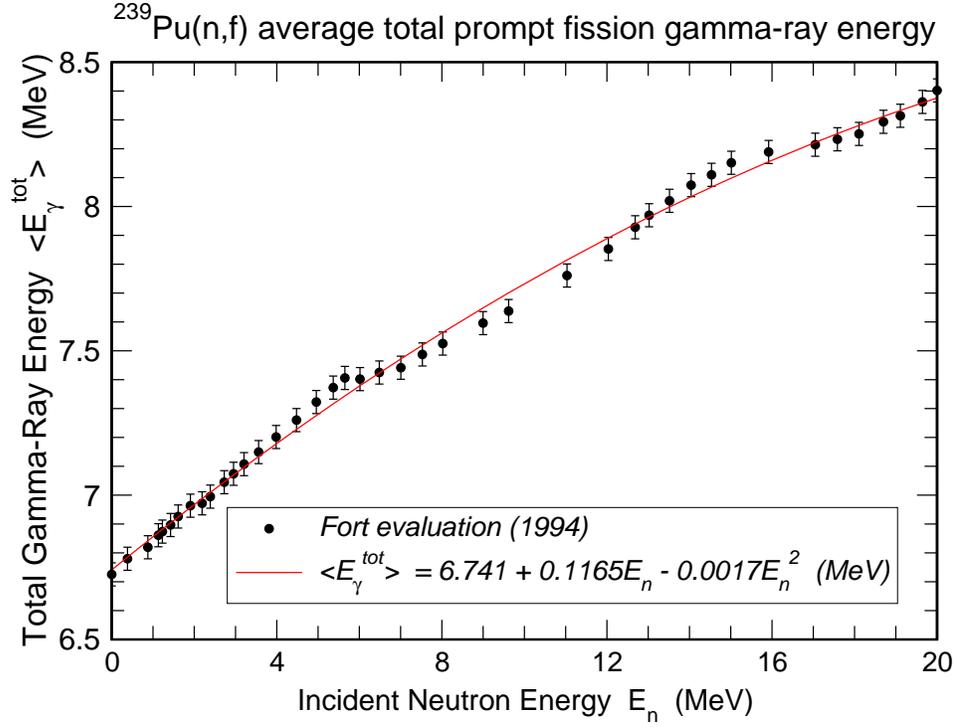, width=3.0in, angle=-90.}
\vspace{90pt}
\caption{Average total prompt fission gamma-ray energy
 $\rm \langle E_{\gamma}^{\rm tot}\rangle$ for the $\rm n(E_{\rm n}$) + $^{239}$Pu system.}
\label{fig16}
\end{figure}

\noindent Consequently, we employ an evaluation by Fort (1994) \cite{For04} which is
based upon systematics with respect to the measurements by Frehaut {\it et al.} \cite{Fr82}
on nearby actinides and upon multichance fission probabilities from the Japanese Nuclear Data
Center. This evaluation is shown as the points appearing in Fig. \ref{fig16} together with a
quadratic fit in the incident neutron energy.
 
\newpage
 
\newpage
 
\noindent {\bf For the n + $^{239}$Pu system:}
 
\begin{eqnarray}
\langle E_{\gamma}^{tot}\rangle &=& (6.741 \pm 0.02) + (0.1165 \pm 0.004)E_{n} \nonumber \\
&-& \mbox{} (0.0017 \pm 0.0002)E_{n}^{2} \; \; (MeV)
\label{p9gam}
\end{eqnarray}
 
\section{Average Total Prompt Fission Energy Release}
 
\noindent Equation (\ref{eqer}) for the average total prompt energy release in fission
can now be evaluated for the three systems under study:
 
\noindent {\bf For the n + $^{235}$U system} one substitutes Eq. (\ref{eq16}), Eq. (\ref{u5ex}), and
Eq. (\ref{u5gam}) into Eq. (\ref{eqer}), together with the value $B_{n}$ = 6.546 MeV for this
system, to obtain
 
\begin{equation}
\langle E_{r}\rangle = 185.6 - 0.0995E_{n} \; \; (MeV)
\label{u5er}
\end{equation}
 
\noindent {\bf For the n + $^{238}$U system} one substitutes Eq. (\ref{eq18}), Eq. (\ref{u8ex}),
and Eq. (\ref{u8gam}) into Eq. (\ref{eqer}), together with the value $B_{n}$ = 4.806 MeV
for this system, to obtain
 
\begin{equation}
\langle E_{r}\rangle = 187.7 - 0.1318E_{n} + 0.0034E_{n}^{2} \; \; (MeV)
\label{u8er}
\end{equation}
 
\noindent {\bf For the n + $^{239}$Pu system} one substitutes Eq. (\ref{eq20}), Eq. (\ref{pu9ex}),
and Eq. (\ref{p9gam}) into Eq. (\ref{eqer}), together with the value $B_{n}$ = 6.534 MeV
for this system, to obtain
 
\begin{equation}
\langle E_{r}\rangle = 197.2 - 0.1617E_{n} - 0.0017E_{n}^{2} \; \; (MeV)
\label{pu9er}
\end{equation}
 
\begin{figure} [b!]
\vspace{-0.5in}
\epsfig{file=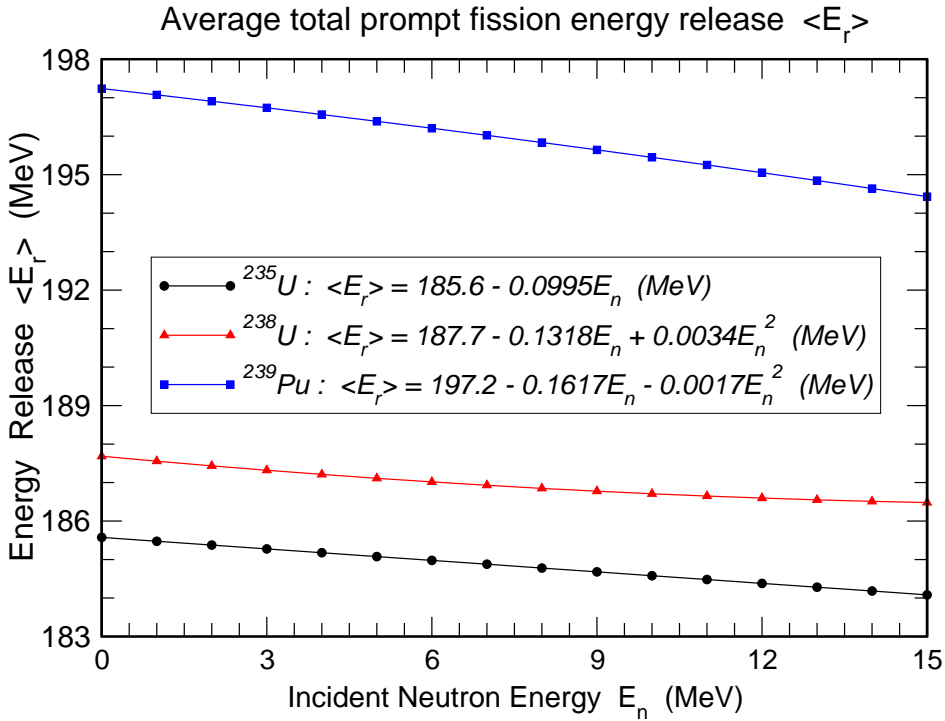, width=3.0in, angle=-90.}
\vspace{90pt}
\caption{Average total prompt fission energy release $\langle \rm E_{\rm r}\rangle$
for three systems.}
\label{fig22}
\end{figure}

\noindent The calculated average total prompt fission energy release
for $\langle E_{r}\rangle$ the three systems is shown in
Fig. \ref{fig22}. There are two major features in this figure. First, the prompt energy
release for the neutron-induced fission of $^{239}$Pu is about 10 MeV greater than
that of the two isotopes of U, over the entire energy range of 15 MeV, and the prompt
energy release for the neutron-induced fission of $^{238}$U is about 2 MeV greater
than that of $^{235}$U, over the same energy range. Second, the prompt energy release
decreases with increasing incident neutron energy for all three of the systems under study,
which is contrary to intuition.
 
\newpage
 
\noindent As already pointed out, this behavior is primarily due to the facts that symmetric
fission increases
with increasing neutron energy and that the total kinetic energy for symmetric fission is
significantly less than the total kinetic energy for the more probable asymmetric fission.
At the same time, however, the average prompt fission neutron multiplicities
as a function of fission-fragment mass, ${\bar{\nu}}_{p}(A_{f})$, are peaked
for symmetric fision.
Note that for far asymmetric fission (which is relatively infrequent) the total
kinetic energy is also significantly reduced.
 
\noindent In the next section we examine the average total prompt energy deposition
in the medium in which
the fission event occurs, for the three systems under study.
 
\section{Average Total Prompt Fission Energy Deposition}
 
\noindent The {\it average total prompt fission energy deposition in the medium for
binary fission  $\langle E_{d}\rangle$
is defined as the average total prompt fission energy release in binary fission
$\langle E_{r}\rangle$
plus the total energy brought to the fission event by the particle inducing
the fission [the quantity ($E_{n} + B_{n}$) for neutron-induced fission] minus
the average total binding energy of the prompt neutrons emitted by the fission
fragments ${\bar{\nu}}_{p}\langle S_{n}\rangle$ [which is not deposited in the medium]}.
With this definition and Eq. (\ref{eqer}) we obtain
 
\begin{eqnarray}
\langle E_{d}\rangle &=& \langle E_{r}\rangle + (E_{n} + B_{n}) - {\bar{\nu}}_{p}\langle S_{n}\rangle \label{eqedo} \\
&=& \langle T_{f}^{tot}\rangle + {\bar{\nu}}_{p}\langle \varepsilon\rangle +
\langle E_{\gamma}^{tot}\rangle \label{eqed} \; .
\end{eqnarray}
 
\noindent We wish to express Eq. (\ref{eqed}) in terms of laboratory observables
  in the medium.
Therefore, note that Eq. (\ref{eq15}) is of the form $\langle T_{p}^{tot}\rangle =
\langle T_{f}^{tot}\rangle[1 - x]$ where $x$ is a small quantity.
Solving Eq. (\ref{eq15}) for $\langle T_{f}^{tot}\rangle$ and performing an expansion of
$1/[1 - x]$ yields
 
\begin{equation}
\langle T_{f}^{tot}\rangle = \langle T_{p}^{tot}\rangle\left[1 + \frac{{\bar{\nu}}_{p}}{2A}
\left(\frac{\langle A_{H}\rangle}{\langle A_{L}\rangle} + \frac{\langle A_{L}\rangle}{\langle A_{H}\rangle}\right)\right]
\label{texp}
\end{equation}
 
\noindent Inserting Eq. (\ref{texp}) into Eq. (\ref{eqed}) and rearranging terms gives
 
\begin{equation}
\langle E_{d}\rangle = \langle T_{p}^{tot}\rangle + \langle E_{neut}^{tot}\rangle + \langle E_{\gamma}^{tot}\rangle
\label{edl}
\end{equation}
 
\noindent with $\langle E_{neut}^{tot}\rangle$ the average total
prompt fission neutron kinetic energy in the laboratory system given by
 
\begin{eqnarray}
\langle E_{neut}^{tot}\rangle &=& {\bar{\nu}}_{p}\left[\frac{1}{2}\left(\frac{\langle A_{H}\rangle}{\langle A_{L}\rangle}
\frac{\langle T_{f}^{tot}\rangle}{A} + \frac{\langle A_{L}\rangle}{\langle A_{H}\rangle}
\frac{\langle T_{f}^{tot}\rangle}{A}\right) + \langle \varepsilon\rangle\right] \label{eneut1} \\
&=& {\bar{\nu}}_{p}\left[\frac{1}{2}(\langle E_{f}^{L}\rangle + \langle E_{f}^{H}\rangle) + \langle \varepsilon\rangle\right] \label{eneut2} \\
&=& {\bar{\nu}}_{p}\langle E\rangle \label{eneut3}
\end{eqnarray}
 
\noindent in which the prompt fission neutron spectrum average laboratory energy
$\langle E\rangle$ is given by the Los Alamos model \cite{MN}
 
\begin{equation}
\langle E\rangle = \frac{1}{2}(\langle E_{f}^{L}\rangle + \langle E_{f}^{H}\rangle) + \langle \varepsilon\rangle \; \; ,
\label{eavg}
\end{equation}
 
\noindent and the approximation has been made in Eq. (\ref{eneut1}) that $\langle T_{p}^{tot}\rangle/A$
can be replaced by $\langle T_{f}^{tot}\rangle/A$ given that the $\langle A_{L,H}\rangle$ are not
unique, but instead averages, and given that $\langle \varepsilon\rangle$ is the dominant term in the
square brackets.
In Eqs. (\ref{eneut2}) and (\ref{eavg}) the quantities $\langle E_{f}^{L}\rangle$ and
$\langle E_{f}^{H}\rangle$ are the average kinetic energies per nucleon of
the moving light and heavy fission fragments, respectively, and are the quantities not included in $\langle Ex_{n}^{tot}\rangle$
as discussed in the beginning of Sec. 3.2.
 
\noindent Note that all averaged quantities appearing in Eqs. (\ref{edl} - \ref{eavg})
depend upon the incident neutron energy $E_{n}$ which has been suppressed for brevity.
We use Eq. (\ref{edl}) for the average total prompt fission energy deposition for the
remainder of this paper.
 
\noindent The evaluation of Eq. (\ref{eneut3}) for the three systems of interest
is performed using the ${\bar{\nu}}_{p}$ values from ENDF \cite{ENDF} shown
in Fig. \ref{fig8} and prompt fission neutron spectrum average laboratory
energies $\langle E\rangle$ calculated for multi-chance fission with the Los
Alamos model \cite{MN} :
 
\begin{equation}
\langle E\rangle = \frac{[P_{f_{1}}^{A}{\bar{\nu}}_{p_{1}}\langle E_{1}\rangle +
P_{f_{2}}^{A}(\langle \xi_{1}\rangle + {\bar{\nu}}_{p_{2}}\langle E_{2}\rangle)
+ P_{f_{3}}^{A}(\langle \xi_{1}\rangle + \langle \xi_{2}\rangle +
{\bar{\nu}}_{p_{3}}\langle E_{3}\rangle)]}
{[P_{f_{1}}^{A}{\bar{\nu}}_{p_{1}} + P_{f_{2}}^{A}(1 + {\bar{\nu}}_{p_{2}}) +
P_{f_{3}}^{A}(2 + {\bar{\nu}}_{p_{3}})]}
\label{emlab}
\end{equation}
 
\noindent where the $\langle E_{i}\rangle$ are the average laboratory neutron
energies for {\it i}th-chance fission and all other quantities are defined as
for Eq. (\ref{epscm}).
 
\begin{figure} [b!]
\vspace{-0.5in}
\epsfig{file=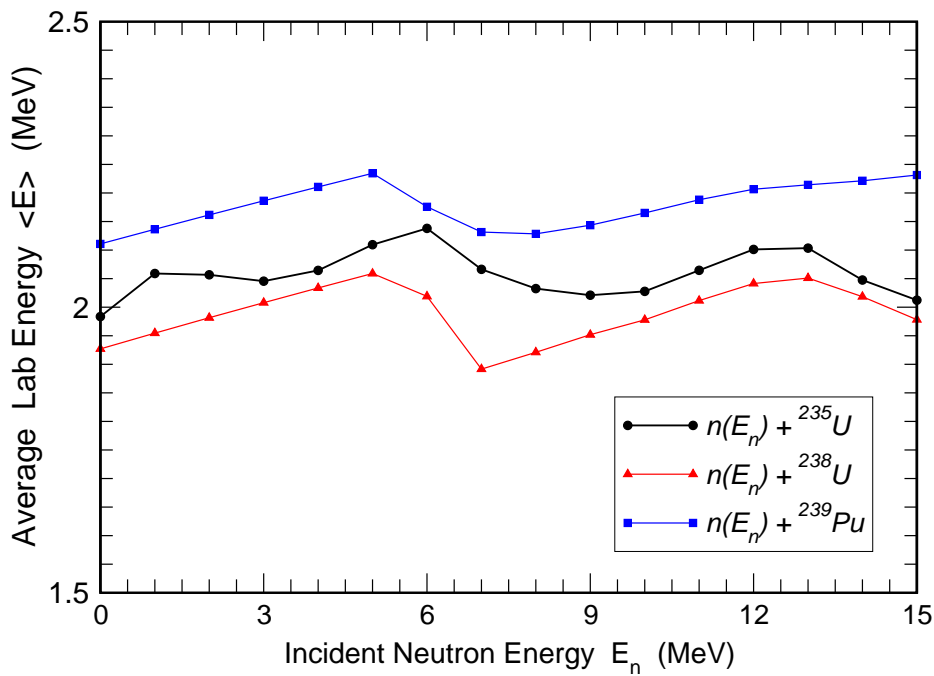,width=2.27in,angle=-90.}
\vspace{130pt}
\caption{Prompt fission neutron spectrum average laboratory energy
$\langle E\rangle$ for three systems calculated with the Los Alamos model (line
segments are to guide the eye).}
\label{fig17}
\end{figure}

\begin{figure} [htb]
\vspace{-0.5in}
\epsfig{file=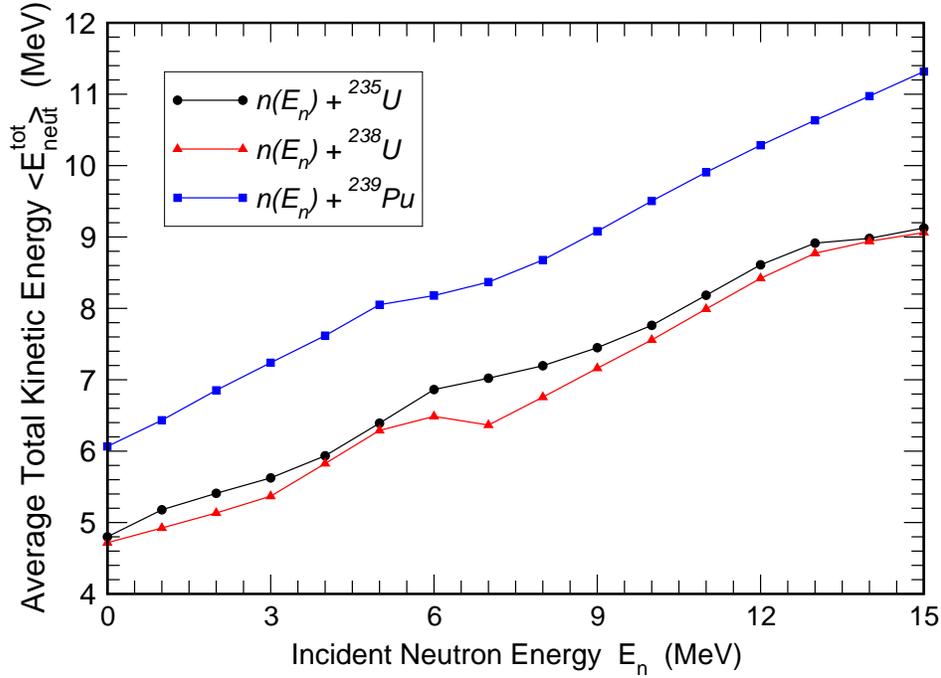, width=2.27in, angle=-90.}
\vspace{130pt}
\caption{Average total prompt fission neutron kinetic energy $\langle \rm E_{\rm neut}^{\rm tot}\rangle$
for three systems from the Los Alamos model and ENDF (line segments are to guide the eye).}
\label{fig18}
\end{figure}

\noindent The prompt fission neutron spectrum
average laboratory energies $\langle E\rangle$ calculated with Eq. (\ref{emlab}) for the three
systems are shown in Fig. \ref{fig17}. Just as with the average center-of-mass
energies $\langle \varepsilon\rangle$, the values of $\langle E\rangle$ decrease
near 6 MeV and 13 MeV, the approximate thresholds for 2nd- and 3rd-chance fission,
and for the same reasons. The product of these two quantities $\langle E_{neut}^{tot}\rangle$
is shown as a function of incident neutron energy in Fig. \ref{fig18}
for the three systems of interest.
Some evidence of multiple-chance fission threshold structure is still present.
 
\begin{figure} [b!]
\vspace{-0.5in}
\epsfig{file=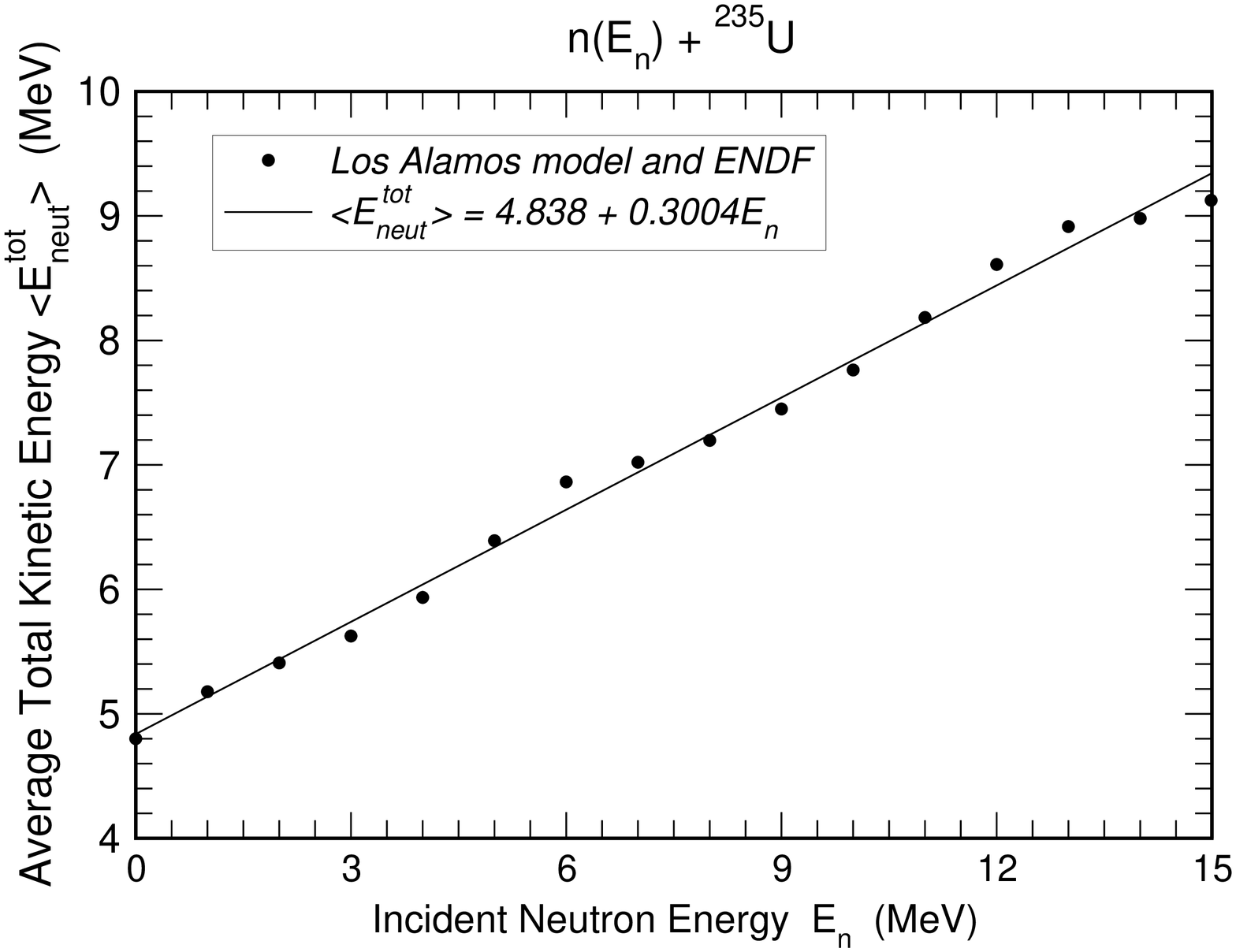, width=2.27in, angle=-90.}
\vspace{130pt}
\caption{Average total prompt fission neutron kinetic energy
 $\langle \rm E_{\rm neut}^{\rm tot}\rangle$ for the n(E$_{\rm n}$) + $^{235}$U system.}
\label{fig19}
\end{figure}

\begin{figure} [htb]
\vspace{-0.5in}
\epsfig{file=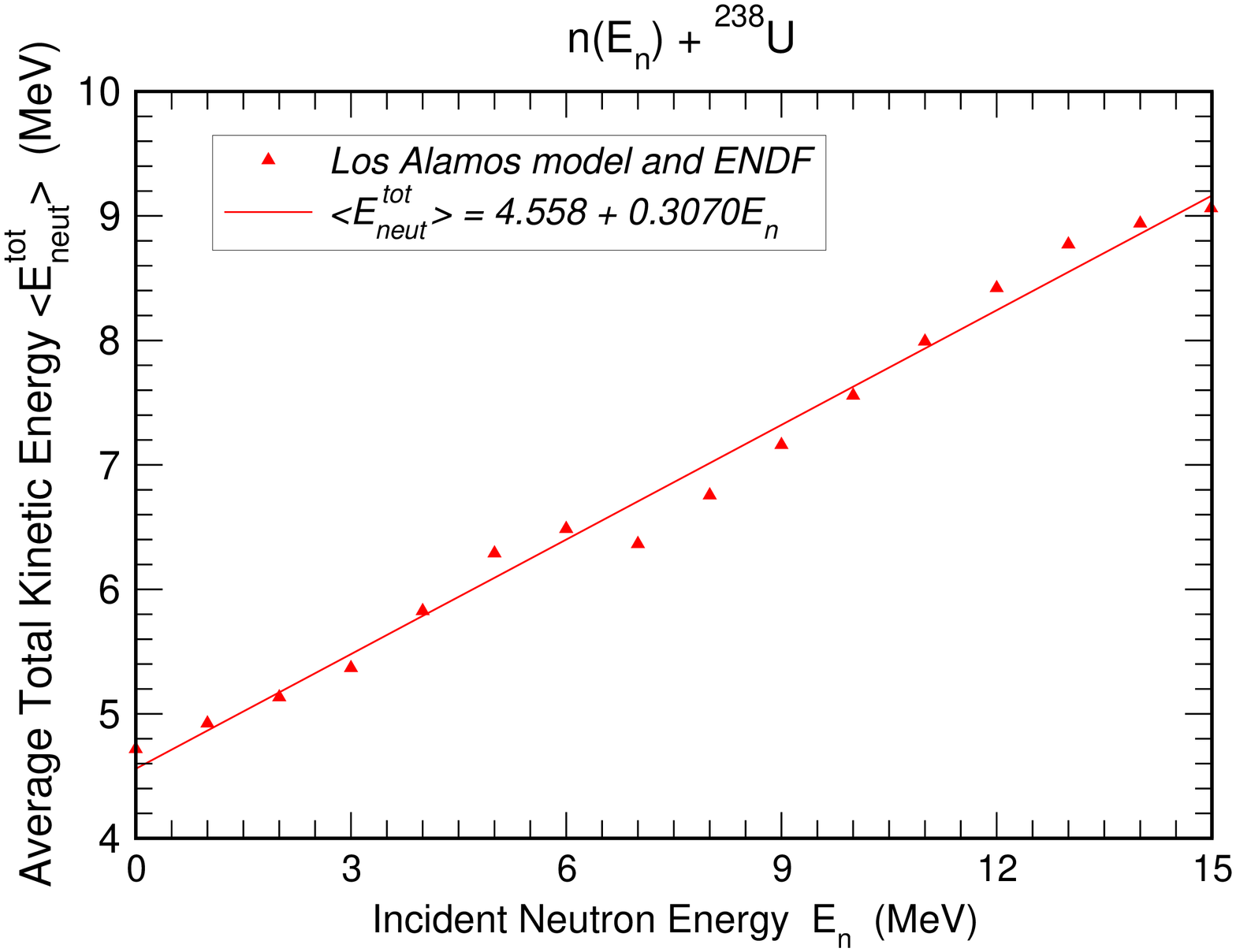, width=2.27in, angle=-90.}
\vspace{130pt}
\caption{Average total prompt fission neutron kinetic energy
 $\langle \rm E_{\rm neut}^{\rm tot}\rangle$ for the n(E$_{\rm n}$) + $^{238}$U system.}
\label{fig20}
\end{figure}

\begin{figure} [b!]
\vspace{-0.5in}
\epsfig{file=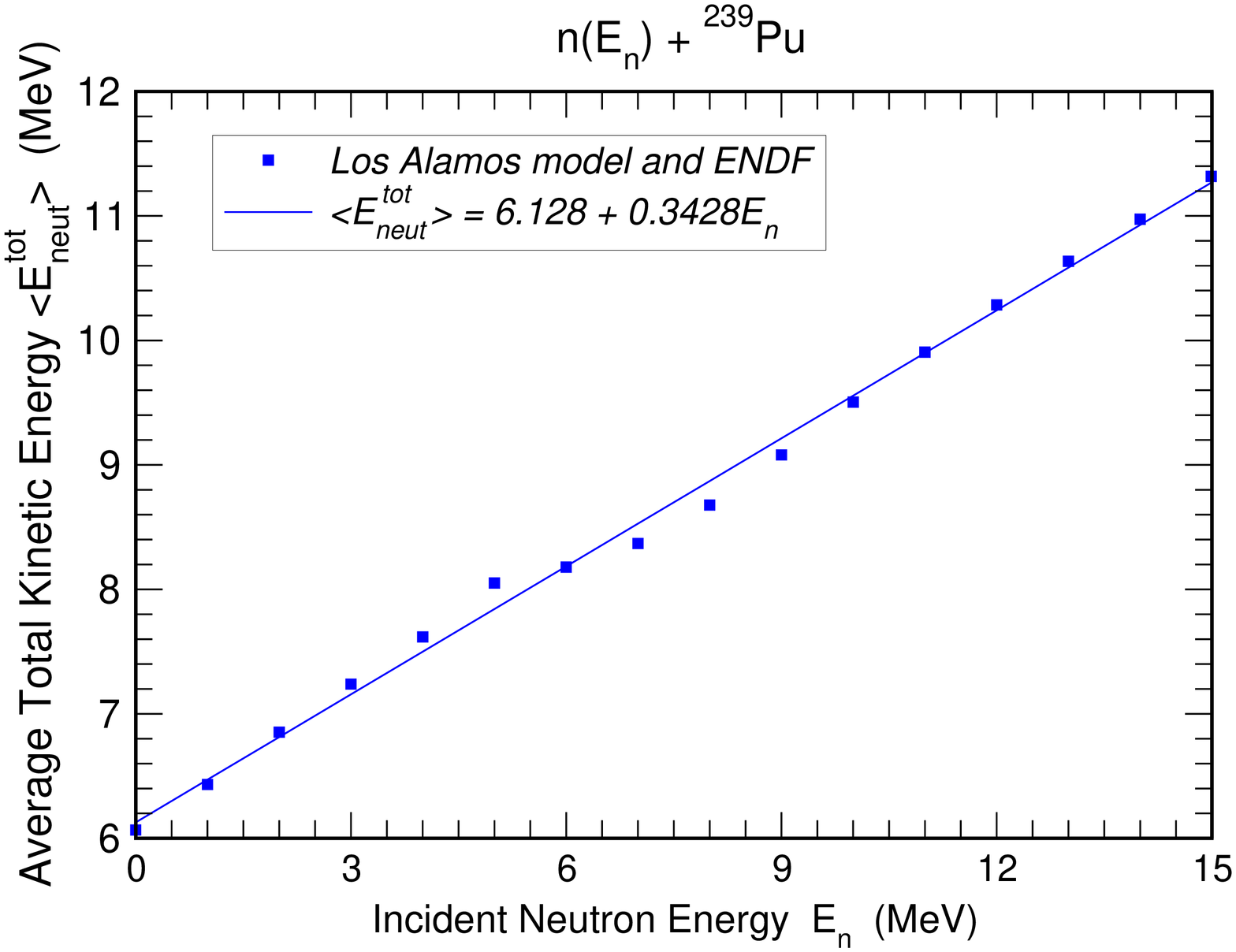, width=2.27in, angle=-90.}
\vspace{130pt}
\caption{Average total prompt fission neutron kinetic energy
 $\langle \rm E_{\rm neut}^{\rm tot}\rangle$ for the n(E$_{\rm n}$) + $^{239}$Pu system.}
\label{fig21}
\end{figure}

\noindent Again, for our present purposes, we perform linear fits to the values
of the average total prompt fission neutron kinetic energies $\langle E_{neut}^{tot}\rangle$
shown in Fig. \ref{fig18} for the three systems. The results are shown
together with the calculated values in Figs. \ref{fig19}, \ref{fig20}, and
\ref{fig21}.
 
 
\noindent {\bf For the n + $^{235}$U system:}
 
\begin{equation}
\langle E_{neut}^{tot}\rangle = 4.838 + 0.3004E_{n} \; \; (MeV)
\label{ent5}
\end{equation}
 
\noindent {\bf For the n + $^{238}$U system:}
 
\begin{equation}
\langle E_{neut}^{tot}\rangle = 4.558 + 0.3070E_{n} \; \; (MeV)
\label{ent8}
\end{equation}
 
\noindent {\bf For the n + $^{239}$Pu system:}
 
\begin{equation}
\langle E_{neut}^{tot}\rangle = 6.128 + 0.3428E_{n} \; \; (MeV)
\label{ent9}
\end{equation}
 
\newpage
 
\newpage
 
\newpage
 
\begin{figure} [htb]
\vspace{-0.5in}
\epsfig{file=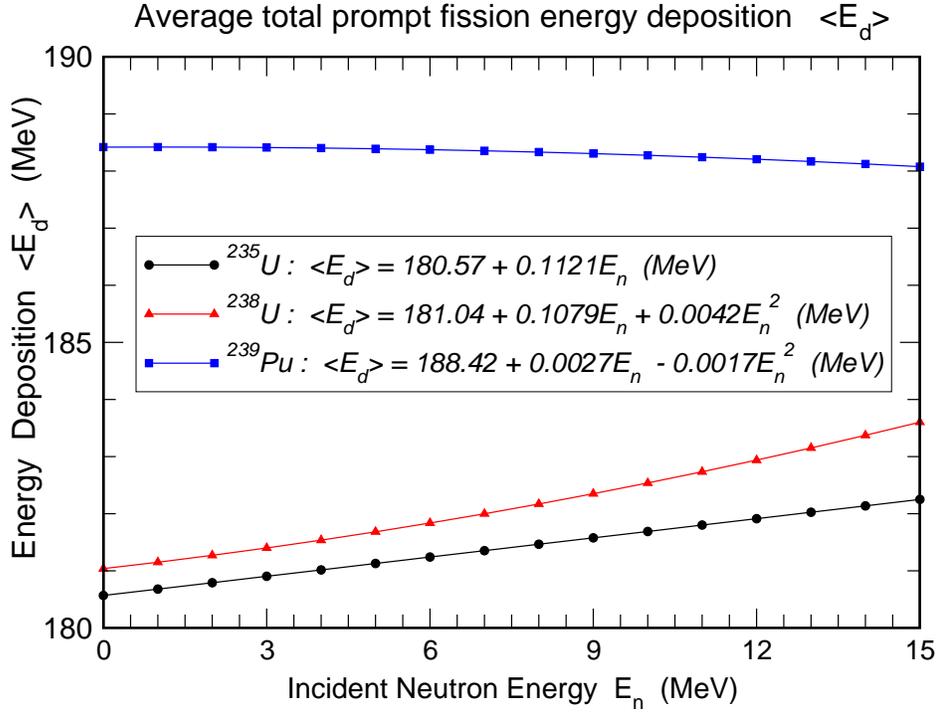, width=2.27in, angle=-90.}
\vspace{130pt}
\caption{Average total prompt fission energy deposition $\langle \rm E_{\rm d}\rangle$
for three systems.}
\label{fig23}
\end{figure}
 
\noindent Equation (\ref{edl}) for the average total prompt fission energy deposition
$\langle E_{d}\rangle$ can now be evaluated for the three systems under study: \\
 
\noindent {\bf For the n + $^{235}$U system} one substitutes Eq. (\ref{eq17}), Eq. (\ref{ent5}),
and Eq. (\ref{u5gam}) into Eq. (\ref{edl}) to obtain
 
\begin{equation}
\langle E_{d}\rangle = 180.57 + 0.1121E_{n} \; \; (MeV)
\label{edu5}
\end{equation}
\noindent {\bf For the n + $^{238}$U system} one substitutes Eq. (\ref{eq19}),
Eq. (\ref{ent8}), and Eq. (\ref{u8gam}) into Eq. (\ref{edl}) to obtain
 
\begin{equation}
\langle E_{d}\rangle = 181.04 + 0.1079E_{n} + 0.0042E_{n}^{2} \; \; (MeV)
\label{edu8}
\end{equation}
 
\noindent {\bf For the n + $^{239}$Pu system} one substitutes Eq. (\ref{eq21}),
Eq. (\ref{ent9}), and Eq. (\ref{p9gam}) into Eq. (\ref{edl}) to obtain
 
\begin{equation}
\langle E_{d}\rangle = 188.42 + 0.0027E_{n} - 0.0017E_{n}^{2} \; \; (MeV) \; .
\label{edpu9}
\end{equation}
 
\noindent These three equations for the average total prompt fission energy deposition
are shown as a function of incident neutron energy
in Fig. \ref{fig23}. The energy dependence of $\langle E_{d}\rangle$ is weak, increasing slightly with incident
neutron energy for the two $U$ isotopes and decreasing very slightly for the $Pu$
isotope. Again, the energy dependence is contrary to intuition and the explanation is the same as
that for the energy release $\langle E_{r}\rangle$. However, the $\langle E_{d}\rangle$ energy
dependence is stronger than that of $\langle E_{r}\rangle$ because the (positive) term
$(E_{n} + B_{n})$ wins over the (negative) term ${\bar{\nu}}_{p}\langle S_{n}\rangle$ in
Eq. (\ref{eqedo}).

\section{Conclusions}
 
\noindent The total prompt fission energy release and energy deposition,
together with their components, have been determined as a function
of the kinetic energy of the neutron inducing the fission, for $^{235}$U, $^{238}$U,
and $^{239}$Pu.
This study has relied primarily upon existing (published) experimental measurements
and secondarily upon nuclear theory and nuclear models.
 
\noindent Contrary to basic physical intuition, it has been found that the
energy release decreases somewhat with incident neutron energy and the energy deposition
changes slightly with incident neutron energy.
The main reason for
this behavior is that symmetric fission increases with increasing incident
neutron energy, but fission-fragment kinetic energies are at a minimum for
symmetric fission. Even more striking is the fact that the Q values for
fission are, on average, somewhat larger for symmetric fission than they
are for asymmetric fission. The extra available fission-fragment excitation
energy at symmetric fission, due to the smaller fragment kinetic energies
and larger fission Q values, results in a peak in the ${\bar{\nu}}_{p}(A_{f})$
{\it vs.} $A_{f}$ curve, that is, the prompt neutron multiplicities are
peaked at symmetric fission.
 
\noindent These results would become more physically correct with a number of more
complete and more accurate measurements over the incident neutron energy range.
Then, instead of the simple linear and quadratic energy dependencies used here,
more realistic characterizations of the incident neutron energy dependencies
could be performed, for example, near the thresholds for multi-chance fission.
In the $n + ^{238}U$ system this could already be done for the average total
fission product kinetic energy $\langle T_{p}^{tot} \rangle$, but is pointless
to do so now because there are no measurements at all of the average total prompt
fission gamma-ray energy $\langle E_{\gamma}^{tot} \rangle$ as a function of incident
neutron energy for this same system.
The current experimental data for fission product kinetic energies in the $n + ^{235}U$ and
$n + ^{239}Pu$ systems allow, at best, linear energy dependencies.
 
\noindent These results would also become more physically correct if the calculations
performed using the Los Alamos model \cite{MN} were replaced by the identical calculations
performed using a modern Hauser-Feshbach approach. Here the competition between neutron and
gamma emission from fission fragments would be treated exactly and the angular momentum
dependencies would be treated for each particle type emission.
However, such an approach does not yet exist because not enough is known to adequately
specify the fission fragment initial conditions across the fragment mass and charge
distributions. Namely, (a) the partition of fissioning compound nucleus excitation energy
between the light and
heavy fragments, (b) the fragment initial angular momenta, and (c) the fragment
initial parities, must all be specified for approximately 400 fragments. In addition,
the fragment mass and charge yield distributions must be known as a function of incident
neutron energy in order to properly weight the approximately 400
Hauser-Feshbach results for each incident neutron energy.
There does not yet exist a calculation of these yield distributions of sufficient accuracy
to perform this task, and measured fission product mass and charge yield distributions
allowing construction of the corresponding fragment distributions are sufficiently complete
at only two incident neutron energies: thermal and 14 MeV.
 
\noindent It is astonishing to find that after some 60 years of fission studies
the post-scission fission observables for the three major actinides $^{235}$U,
$^{238}$U, and $^{239}$Pu have been so incompletely measured and understood.
As a consequence of this work the following measurements are recommended for
the three systems studied:
 
\begin{center}
{\bf n + $^{235}$U}
\end{center}
\begin{description}
\item[(a)] The average total fission-product kinetic energy should be measured from 10 keV to 30 MeV.
The existing data stops at about 9 MeV and, furthermore, has uncertainties
that need to be reduced.
\item[(b)] The average total prompt fission gamma-ray energy should be measured
from 15 MeV to 30 MeV and several (three) of Frehaut's data points between
1 and 15 MeV should be remeasured  to verify the energy dependence.
\item[(c)] The prompt fission neutron spectrum (out to 15 MeV emitted neutron energy)
should be remeasured for incident neutron energies of 1, 2, 3, and 5 MeV,
and measured for 4 MeV and the range 8 MeV to 30 MeV.
\end{description}
 
\begin{center}
{\bf n + $^{238}$U}
\end{center}
\begin{description}
\item[(a)] The average total prompt fission gamma-ray energy should be measured
from 10 keV to 30 MeV. Apparently, no measurements exist.
\item[(b)] The prompt fission neutron spectrum (out to 15 MeV emitted neutron energy)
should be measured for incident neutron energies of 4, 10, and the range 12 to 30 MeV.
\end{description}
 
\begin{center}
{\bf n + $^{239}$Pu}
\end{center}
\begin{description}
\item[(a)] The average total fission-product kinetic energy should be measured from about
3 MeV to 30 MeV.
The existing measurements become sparse at about 3.5 MeV and stop just beyond 5 MeV.
\item[(b)] The average total prompt fission gamma-ray energy should be measured
from 10 keV to 30 MeV.  Only thermal measurements exist at the present time
and the experimental content of Fort's evaluation is unknown.
\item[(c)] The prompt fission neutron spectrum (out to 15 MeV emitted neutron energy)
should be measured for incident neutron energies ranging from 4 MeV to 30 MeV.
\end{description}
 
\noindent It is manifestly clear that measured components of the above quantities
for the light and heavy mass peaks as well as for near symmetric and far asymmetric fission
will be doubly useful in serving both fundamental and applied post-scission fission physics.
 
\begin{center}
{\bf{Appendix A \ Multiple-Chance Fission Equations}}
\end{center}
 
\noindent In the absence of the measured components of the average total prompt
fission energy release $\langle E_{r}\rangle$ and average total prompt fission
energy deposition $\langle E_{d}\rangle$ the following equations are used to
calculate these quantities directly:
 
\begin{equation}
\langle E_{r}\rangle = \frac{[P_{f_{1}}^{A}\;\langle E_{r}(A)\rangle + P_{f_{2}}^{A}\;\langle
E_{r}(A-1)\rangle + P_{f_{3}}^{A}\;\langle E_{r}(A-2)\rangle]}
{[P_{f_{1}}^{A} + P_{f_{2}}^{A} + P_{f_{3}}^{A}]}
\end{equation}
 
\begin{equation}
\langle E_{d}\rangle = \frac{[P_{f_{1}}^{A}\;\langle E_{d}(A)\rangle + P_{f_{2}}^{A}\;\langle
E_{d}(A-1)\rangle + P_{f_{3}}^{A}\;\langle E_{d}(A-2)\rangle]}
{[P_{f_{1}}^{A} + P_{f_{2}}^{A} + P_{f_{3}}^{A}]}
\end{equation}
 
\noindent where the total fission probability $P_{f}^{A}$ of the compound
fissioning nucleus $A$ at excitation energy $[E_{n} + B_{n}(A)]$ is given by
 
\begin{equation}
P_{f}^{A}[E_{n} + B_{n}(A)] = P_{f_{1}}^{A}[E_{n} + B_{n}(A)] + P_{f_{2}}^{A}
[E_{n} + B_{n}(A)] + P_{f_{3}}^{A}[E_{n} + B_{n}(A)]
\end{equation}
 
\noindent in which $P_{f_{1}}^{A}$ is the probability for first-chance fission,
the $(n,f)$ reaction, $P_{f_{2}}^{A}$ is the probability for second-chance fission,
the $(n,n^{\prime} f)$ reaction, and $P_{f_{3}}^{A}$ is the probability for
third-chance fission, the $(n,n^{\prime} n^{\prime\prime} f)$ reaction.
 
\noindent Correspondingly,
 
\begin{equation}
\langle E_{r}(A)\rangle = \langle T_{f}^{tot}(A)\rangle + {\bar{\nu}}_{p}(A)[\langle
S_{n}(A)\rangle + \langle \epsilon_{1}(A)\rangle] + \langle E_{\gamma}^{tot}(A)\rangle
- [E_{n} + B_{n}(A)]
\end{equation}
 
\begin{equation}
\langle E_{d}(A)\rangle = \langle T_{p}^{tot}(A)\rangle + {\bar{\nu}}_{p}(A)\langle E(A)\rangle
+ \langle E_{\gamma}^{tot}(A)\rangle
\end{equation}
 
\noindent which are evaluated as a function of excitation energy $[E_{n} + B_{n}(A)]$ of the
$A$ system,
 
\begin{eqnarray}
\langle E_{r}(A-1)\rangle & = & \langle T_{f}^{tot}(A-1)\rangle + {\bar{\nu}}_{p}(A-1)
[\langle S_{n}(A-1)\rangle + \langle \epsilon_{2}(A-1)\rangle]
\nonumber \\
&+& \langle E_{\gamma}^{tot}(A-1)\rangle - [E_{n} - \langle \xi_{1}(A)\rangle]
\end{eqnarray}
 
\begin{equation}
\langle E_{d}(A-1)\rangle = \langle T_{p}^{tot}(A-1)\rangle + {\bar{\nu}}_{p}(A-1)
\langle E(A-1)\rangle + \langle E_{\gamma}^{tot}(A-1)\rangle
\end{equation}
 
\noindent which are evaluated as a function of excitation energy $[E_{n} - \langle \xi_{1}(A)\rangle]$
of the $A-1$ system, and
 
\begin{eqnarray}
\langle E_{r}(A-2)\rangle & = & \langle T_{f}^{tot}(A-2)\rangle + {\bar{\nu}}_{p}(A-2)
[\langle S_{n}(A-2)\rangle + \langle \epsilon_{3}(A-2)\rangle]
\nonumber \\
&+& \langle E_{\gamma}^{tot}(A-2)\rangle - [E_{n} - \langle \xi_{1}(A)\rangle - \langle \xi_{2}(A-1)\rangle
\nonumber \\
&-& B_{n}(A-1)]
\end{eqnarray}
 
\begin{equation}
\langle E_{d}(A-2)\rangle = \langle T_{p}^{tot}(A-2)\rangle + {\bar{\nu}}_{p}(A-2)\langle E(A-2)\rangle
+ \langle E_{\gamma}^{tot}(A-2)\rangle
\end{equation}
 
\noindent which are evaluated as a function of excitation energy $[E_{n}
- \langle \xi_{1}(A)\rangle - \langle \xi_{2}(A-1)\rangle - B_{n}(A-1)]$ of
the $A-2$ system. \\
 
\noindent All quantities appearing in the equations of this appendix are defined
elsewhere in the text, mostly near Eq. (\ref{epscm}) and Eqs. (\ref{eavg} - \ref{emlab}).


\begin{thebibliography}{99}
\bibitem{note}
Note that the upper limit, $\sim 10^{-7}$ [sec], may be application dependent and could,
for example, be replaced by $\sim 10^{-8}$ or $\sim 10^{-6}$ [sec].
\end{thebibliography}
\end{document}